\documentclass[useAMS,usenatbib]{mn2e}
\usepackage[english]{babel}        
\usepackage{graphicx}
\usepackage{rotating}
\usepackage{float}
\title[SNe Ia calibration \& $H_0$]
{Cepheid calibration of type Ia Supernovae and the Hubble constant
}

\author[Altavilla et al.]  {G. Altavilla$^{1,2}$\thanks{E-mail:
altavilla@pd.astro.it}, G. Fiorentino$^3$, M. Marconi$^3$, I. Musella$^3$, 
E. Cappellaro$^3$, \and R. Barbon$^{2}$, S. Benetti$^{1}$,
A. Pastorello$^{1,2}$, M. Riello$^{1,2}$,  M.~Turatto$^{1}$, L.~Zampieri$^{1}$ 
\\
$^1$INAF - Padova Astronomical Observatory, vicolo dell'Osservatorio 5,
I-35122 Padova, Italy \\
$^2$Department of Astronomy, Padova University, Vicolo dell'Osservatorio 2, I-35122 Padova, 
Italy\\
$^3$INAF - Capodimonte Astronomical Observatory, Via Moiariello 16, 
I-80131 Napoli, Italy\\
}
\begin{document}

\date{Received ................; accepted ................}

%\pagerange{\pageref{firstpage}--\pageref{lastpage}} \pubyear{2002}

\maketitle
               
\begin{abstract}

We investigate how a different calibration of the Cepheid
Period-Luminosity (PL) relation taking into account the metallicity
corrections, affects the absolute magnitude calibration of Supernovae
(SNe) Ia and, in turn, the determination of the Hubble constant $H_0$.
We exploit  SN Ia light curves from literature and previously unpublished
data, to build the $M_B$ -- $\Delta m_{15}(B)$ relation and we calibrate the
zero point by means of 9 type Ia SNe with Cepheid measured distances.
This relation was then used to build the Hubble diagram and in turn to
derive $H_0$. In the attempt to correct for the host galaxy extinction,
we found that the data seems to suggest a value for the total to
selective absorption ratio, $R_B=3.5$, which is smaller than the standard
value for our own Galaxy $R_B=4.315$. 

Depending on different metallicity corrections for the Cepheids P-L
relation, values of $R_B$ and SN sample selection criteria, we found
that the values of the Hubble constant $H_0$ is in the range 68--74
\,km\,s$^{-1}$\,Mpc$^{-1}$, with associated uncertainties of the order of
$10\%$.

Unpublished photometry is also presented
for 18 SNe of our sample 
(1991S, 1991T, 1992A, 1992K, 1993H, 1993L, 1994D, 1994M, 1994ae, 
1995D, 1995ac,  1995bd,  1996bo, 1997bp, 1997br, 1999aa, 1999dk, 
2000cx), which are  the results of a 
long standing effort for supernova monitoring at ESO - La Silla and Asiago
Observatories.

\end{abstract} 
 

\section{Introduction} \label{int}

Type Ia supernovae are probably the most accurate distance indicators
on cosmological scales. An important limitation is that, since SNe are
relatively rare events, they cannot be used to measure the distance of
pre-selected individual galaxies. On the other side they are
invaluable tools to measure global properties of our Universe, like
the Hubble constant $H_0$
(e.g. \citealt{rust,hamuy96a,tripp,freedman2001}), the cosmological
parameters $q_0$, $\Omega_m$, $\Omega_{\Lambda}$
(e.g. \citealt{hzss,scp,riess98}), and also the peculiar velocity
field \citep{riess97}.  This explains the continuous efforts spent to
refine the absolute calibration of SNe~Ia.

Three main issues need to be addressed for a proper calibration of
SNe~Ia, namely the direct calibration of individual events through
primary distance indicators, the determination of the relation between
light curve shape and absolute magnitude and the extinction estimate,
in particular that due to the host galaxy dust. This process is
complicated by the fact that these three items have a deep
interplay. This means that it is not easy to predict how the effect of
a new finding on one ingredient propagates to the final
result. Indeed, different assumptions in the calibration chain may
explain why different authors, using the same data, obtain different
results \citep{bruno}.

The purpose of this work is to investigate how new calibrations of the
Cepheids absolute magnitude, discussed in \S \ref{cefeidisection},
propagate through the calibration of SNe~Ia on the determination of
$H_0$. In this work we do not rely on published values of SN light
curve parameters (in particular maximum magnitudes and decline rates),
but we made our own independent estimates.  To this aim we exploited
the archive of SN~Ia light curves which has been collected at the
Padova Observatory. Besides all relevant data published in the
literature (with a major contribution given by \citealt{hamuy96b} and
\citealt{riess99}), the archive includes also the results of a long
standing effort for supernova monitoring at ESO and Asiago
Observatories \citep{turatto00}.
The  unpublished data are 
reported in Appendix~\ref{fotometriapadova}.
\\
After rejecting objects with
incomplete photometric coverage we retained a sample of 78 SNe~Ia (see
Appendix~\ref{Asnedata}).

The plan of the paper is the following: in \S \ref{sniacalibration} we
discuss the SNe Ia inhomogeneity, the properties of their light curves
and the calibration of the absolute magnitude at maximum. In
particular we address the problem of the host galaxy absorption and
the effect of the metal content on the Cepheid distance scale.  In \S
\ref{hubble} we show how different assumptions on the peak magnitude
calibration and different metallicity corrections on the Cepheid PL
relation affect the Hubble constant.  Conclusions are drawn in \S
\ref{conclusion}.

\section{SN\protect\lowercase{e} I\protect\lowercase{a} calibration} \label{sniacalibration}

Contrary to early claims, the absolute magnitude of SNe~Ia at maximum
($M_{max}$) is not constant but it ranges over 2--2.5 mag.  However a
correlation between $M_{max}$ and the shape of the light curve has
been suggested which allows to recover SNe Ia as accurate distance
indicators.  This was first proposed almost thirty years ago
\citep{bcr,rust,pskovskii} but, because of the large photometric
errors of photographic photometry and the contamination of the early
sample from other SN types (SNe Ib/c in particular), remained debated
until a few years ago \citep{sandage}.  Eventually, the improved
accuracy obtained with CCD photometry definitively proved that
$M_{max}$ is brighter in SNe~Ia with a slower luminosity decline
\citep{phil93}.  It is still debated what may be the most convenient
parameter(s) to describe the light curve shape and discriminate
between intrinsically brighter and fainter SNe and, more important,
what is the correct absolute calibration.

Theoretical calculations have shown that the absolute magnitude at
maximum is proportional to the amount of $^{56}$Ni
\citep{arnett,hoflich96,cappellaro97}. However it is not yet clear how
this relates with the progenitor scenarios and explosion mechanism.
In particular, a growing number of ``outliers'' seems to indicate that
a one-parameter description of SNe Ia is not sufficient to
parameterize the light curve of SNe Ia (for an analysis of the type Ia
luminosity dependence on second parameters see \citealt{parodi00}).  In
addition, recent observations show an increasing number of
``disturbing'' objects which merge properties of normal, sub-luminous
and super-luminous SNe, like SN 2000cx or SN 2002cx.  The first one is
characterized by a slow decline but a relatively fast pre-maximum
brightening, similar to that of the normal SN 1994D \citep{weidong01}.
Nevertheless the absolute magnitude at maximum is consistent with a
sub-luminous SN \citep{candia02} and, in addition, the $B-V$ colour
shows an unusual evolution, becoming at late time significantly bluer
than the average of unreddened normal SNe~Ia \citep{candia02}.  The
second one is characterized by a 1991T-like pre-maximum spectrum and a
1991bg-like luminosity, a normal $B-V$ colour evolution, very low
expansion velocities and many other peculiarities \citep{weidong03}.

Therefore, even if we will exploit the fact that in general the light
curve shape correlates with the absolute magnitude at maximum, it
should be kept in mind that this is not a universal law. Not only it
has an intrinsic dispersion but in addition there are really
``anomalous'' SNe.

\subsection{Light curve shape}\label{analysis}

A few different approaches are currently in use to characterize the
luminosity evolution of different SNe~Ia. In principle the Multi-colour
Light Curve Shape (MLCS) method of \cite{riess96a}, which makes use of
the entire early light curve (from discovery to 80-100 days) in
different colours, is expected to be the most robust from a
statistically point of view. This method has also the great advantage
of giving at the same time the absolute calibration and the
extinction.  However the MLCS method heavily relies on the accurate
calibration of a few template light curves, especially for what
concerns the extinction correction, which explains the difference in
the template colour curves in \citealt{riess96a} and \citealt{riess98}
(see also \citealt{salvo}).

\begin{figure}
\begin{center}
\includegraphics[angle=0,width=84mm]{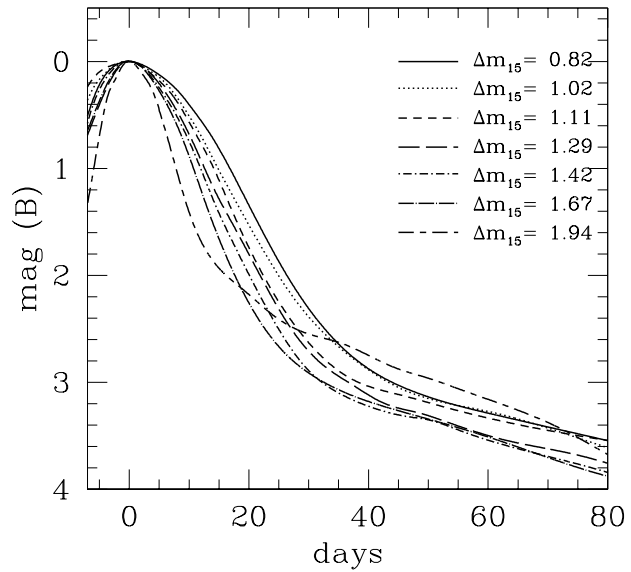}
\caption{SN~Ia template light curves in B band for seven different
decline rates.}
\label{template}
\end{center}
\end{figure}

Two more straightforward measurements of the luminosity evolution are
the $\Delta m_{15}$  (the magnitude decline between the maximum and fifteen days later)
\citep{phil93}, and the ``stretch factor'' $s$ \citep{perlm}, a
coefficient used to expand or contract linearly the time-scale of a
particular light curve in order to match a template\footnote{Recently
\cite{wang2003} introduced a new method called the Colour-Magnitude
Intercept Calibration (CMAGIC) which makes use of multi-colour
post-maximum light curves. It is based on the empirical relation found
between the B magnitude and the $B-V$ (or $B-R$ or $B-I$) colour during the
first month past maximum.  A complete analysis of this method,
discussing also the B-R and B-I colours will be presented by the same
authors in a later paper.}.
In this work we choose to compare these last two methods.

\begin{figure}
\begin{center}
\includegraphics[width=84mm]{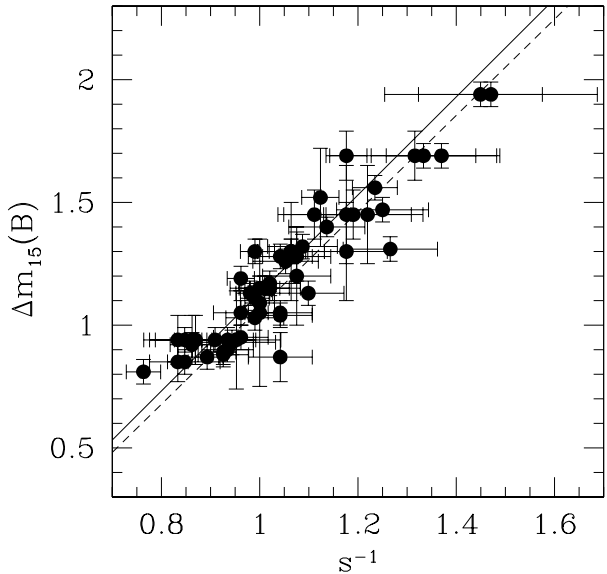} %dm15_s.m 
\caption{$\Delta m_{15}(B)$ versus {\it stretch factor s} for 58
SNe~Ia.  The inferred correlation (solid line) is in good agreement
with that obtained by \citet{perlm} (dashed line). } %macro dm15_s.m
\label{dm15stretch}
\end{center}
\end{figure}

 To measure $\Delta m_{15}(B)$ and $s$ for the SNe of our sample we
proceeded as follows:

\begin{enumerate}

\item B and V light curves of all SNe were corrected for the effect of
red-shift on the time scale of evolution (time dilatation) and on the
photometry (K-correction). For the latter we use the K corrections
given by \cite{hamuy93};

\item among the SNe of our sample we selected 28 events for which the
most complete temporal coverage was available.  For each individual
object we have measured the epoch and magnitude of maximum and the
$\Delta m_{15}(B)$.  The SN light curves were then grouped in seven
bins of similar $\Delta m_{15}(B)$ (in order to have a good $\Delta
m_{15}(B)$ sampling and enough statistics per bin).  The individual
observations were merged together to derive seven template light
curves characterized by the average $\Delta m_{15}(B)$ (see 
Appendix~\ref{templateBnewsec}).  
 Due to the paucity of light curves
with  coverage during the rise phase, the light curve trend
before maximum is quite uncertain.
The  template light curves are
shown in Fig.~\ref{template} where it can be seen that with the
increasing of $\Delta m_{15}(B)$ the light curve shape varies with
continuity. That is, the magnitude differences which originates from
the different early decline rates are more or less maintained after
the inflection point when the light curve turn on a more gentle
decline. The exception are the SNe with the highest early decline
($\Delta m_{15}(B)=1.94$) that, because of an earlier occurrence of the
inflection point, 1-2 months after maximum appear brighter (relatively
to maximum) than other SNe~Ia.

\item for SNe with incomplete temporal coverage we have selected, by
means of a $\chi^2$ test, the best matching template and attributed to
the object the corresponding $\Delta m_{15}(B)$.
For practical reasons, the section of the light curve which goes from 
maximum to the inflection point is that for which detailed monitoring for 
the largest sample of events is available. Therefore, in the light curve 
fit, we gave more weight to the points in this phase range.
Eventually, the
shifts in phase and magnitude adopted to get the best match were used
to refine the initial guess for the epoch and magnitude at maximum.
\end{enumerate}

Instead, to measure the stretch factor $s$, we have adopted as
reference the template light curve with $\Delta m_{15}(B)=1.11$,
i.e. our template closer to the ``standard light curve'', arbitrary
chosen to be characterized by $\Delta m_{15}(B)=1.1$ \citep{perlm}.
In practice, for each SN the time scale of the light curve was
compressed by a factor $s$ chosen to get the best $\chi^2$ fit with
the reference template light curve. 
 Due to the 
uncertainties on the rising branch of the light curve templates and on the 
paucity of observations before maximum, the comparison has been
performed mainly taking into account the light curve evolution from maximum
to 80 days at most.

Following this procedure, we could measure both $\Delta m_{15}(B)$ and
$s$ for 58 SNe Ia (columns 3 and 4 in Table~\ref{tab11}).
  Fig.~\ref{dm15stretch} shows that $\Delta
m_{15}(B)$ and $s^{-1}$ are indeed well correlated.  
We notice that,
as expected from Fig.~\ref{template}, 
the stretched template light curve gives a
very poor fit of the observed light curve for fast declining objects
(like SN 1991bg),
which can not be well fitted simultaneously in  the early and late phases.
This is the reason of the large error-bars for such
objects.  In any case, a least-squares fit considering the errors on
both variables gives $\Delta m_{15}(B)=(1.98\pm 0.16)(s^{-1}-1)+ (1.13
\pm 0.02)$ which is very close to the relation obtained by
\citet{perlm} ($\Delta m_{15}(B)=(1.96\pm 0.17)(s^{-1}-1)+1.07$) from
18 low red-shift SNe.

We also note that the dispersion along the fitting line is consistent
with the error estimates and, even in this enlarged sample, there are
no obvious outliers.  Hence, the systematic differences in the SN
magnitudes calibrated using either $\Delta m_{15}$ or the stretch
factor \citep{drell, bruno} are not due to a misalignment of the
different methods of measuring the light curve shapes and instead must
originate in some other steps of the calibration chain. As we will see
the reddening correction is a most critical issue.

Because of the good correlation between $\Delta m_{15}$ and stretch
factor and because $\Delta m_{15}$ is available for a larger number of
objects, for the SNe~Ia calibration hereafter we will use the $\Delta
m_{15}$ light curve parameterization.

\subsection{Reddening correction}\label{Reddening}

\begin{figure}
\begin{center}
\includegraphics[angle=0,width=84mm]{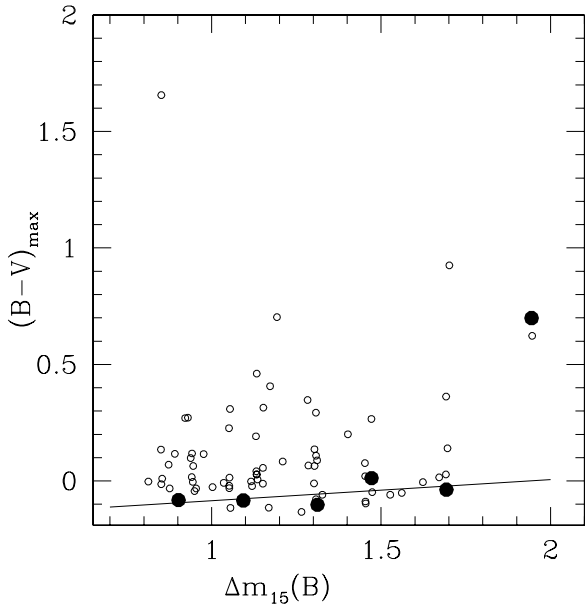} %BVlira2.m
\caption{$(B-V)_{max}$ vs $\Delta m_{15}$.  Correction for Galactic
extinction has been applied.  Black dots are SNe not affected by host
galaxy reddening and spanning a wide range of decline rates. The solid
line is the linear fit of these unreddened SNe (excluding the points
with $\Delta m_{15}\sim 1.95$; see text).  The scatter of the empty
circles is due to the host galaxy reddening.  }
\label{dm15BVmax}
\end{center}
\end{figure}

The SN light is extinguished by the dust in our own Galaxy and in the
host galaxy. For the Galactic extinction we have independent estimates
either from HI maps and galaxy counts \citep{burstein} or from the
COBE/DIRBE and IRAS/ISSA maps \citep*{schlegel}. The latter has been
adopted in the present work.

Instead, to estimate the extinction in the host galaxy, following the
usual approach we make use of the observed SN colour. In practice, we
derive two different estimates as follows:

\begin{enumerate}
\item a plot of the $B-V$ colour at maximum (corrected for galactic
extinction) versus the light curve decline rate $\Delta m_{15}(B)$ is
shown in Fig.~\ref{dm15BVmax}.  The $B-V$ colour at maximum has been
measured on the $B-V$ colour curve by a polynomial fit of the points
around the maximum or by a template colour curve when the sampling was
poor.  We used different templates corresponding to SNe~Ia with
different $\Delta m_{15}$. 

Most of the observed dispersion is due to host galaxy extinction, but
there are also intrinsic differences. In particular it is now known
that SN Ia with $\Delta m_{15}(B)>1.8$ are intrinsically red at
maximum, with $(B-V)_0 \simeq 0.7$. For all other SNe~Ia and assuming
that the lower rim of the distribution corresponds to SNe with
negligible extinction, the differences in the intrinsic colours are
certainly smaller but still there is an indication that in the
intrinsic colour correlates with the luminosity decline rate: SNe with
a faster decline rate are redder \citep{riess98,phil99,nobili2003}.
This is shown in Fig.~\ref{dm15BVmax} by fitting the data of 5 SNe
(SN 1999bc, 1992al, 1994D, 1992A, 1992bo) spanning a wide range of
decline rates and for which there are indications that the host galaxy
reddening was very low \citep{phil99}. We found $(B-V)_{max}=0.09(\pm
0.08)\times (\Delta m_{15}-1.1)-0.08(\pm 0.03)$ and this relation was
adopted to derive the colour excess for all type Ia SNe but the very
fast declining objects ($\Delta m_{15}(B)>1.8$). For the latter we
assumed that they have all the same intrinsic colour $(B-V)_0 = 0.70$.
$E(B-V)_{max}$, corrected for the galactic component, are reported in
column 6 of Table~\ref{tab11}.

\item
As shown by \cite{lira}, the colour evolution of SN~Ia between 30 and
90 days can be well approximated by the following linear relation:
$(B-V)_0=0.725-0.0118 \times (t_V-60)$, where $t_V$ is the time (in
days) from the V maximum. This is independent on the SN decline rates,
including extreme objects like SN~1991bg. There is however at least
one important exception, SN 2000cx \citep{candia02} which should not
be overlooked.

The shift of observed colour of SN~Ia required to fit this relation gives
an alternative estimate of the host galaxy extinction $E(B-V)_{tail}$
reported in column 7 of Table~\ref{tab11}.
\end{enumerate}
\begin{figure}
\begin{center}
\includegraphics[width=84mm]{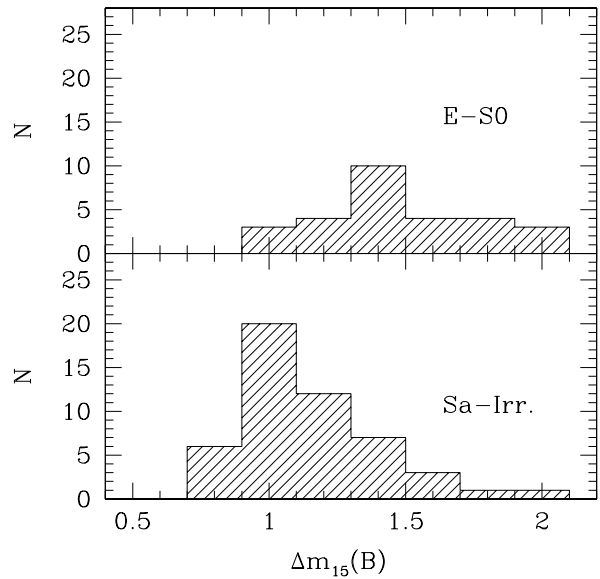} 
%dm15_tipomorfologico.m +KM.m
\caption{Histogram of $\Delta m_{15}(B)$ for SNe Ia in early type
galaxies (upper panel) and in late galaxies (lower panel).  The
$\Delta m_{15}(B)$ values have been corrected for extinction, as
explained in \S \ref{Reddening}.  A Kolmogorof--Smirnov test shows
that the probability that the two distributions derive from the same
population is very small ($P=0.003$\%).   }
\label{dm15tipogalassia}
\end{center}
\end{figure}

Finally, as estimate of the host galaxy extinction, $E(B-V)_{host}$, was
adopted the weighted mean of $E(B-V)_{max}$ and $E(B-V)_{tail}$
(cf. \citealt{phil99}). In the few cases, when the formal mean turns out
to be negative, the $E(B-V)_{host}$ were set equal to zero.
This is equivalent to adopt a Bayesian filter with a flat a priori
distribution for positive $E(B-V)$ and zero for $E(B-V)<0$.
Our adopted  $E(B-V)_{host}$ are reported in column 8 of Table~\ref{tab11}.

Some authors applied a different Bayesian filter to the measured
values of $E(B-V)$. In particular, \cite{phil99} assumed a one-sided
Gaussian ``a priori distribution'' \citep{riess96a}. It turns out that
their choice of a relatively small value of $\sigma$ corresponds to an
arbitrary reduction of $E(B-V)_{host}$, especially for objects with high
colour excess. In turn, this results in smaller absorption corrections
for the SN magnitudes.  This choice, while not well justified, has the
positive effect to reduce the dispersion of the corrected SN absolute
magnitudes. We will come back to this point in the next section.

\cite{phil99} have shown that, because of the rapid spectral evolution
of SN~Ia and of the (small) dependence of the reddening on the colour
of the source the observed $\Delta m_{15}(B)$ must be corrected as
follows: $\Delta m_{15}(B) \simeq \Delta m_{15}(B)_{obs}+0.1 \times
E(B-V)$. The $\Delta m_{15}(B)$ reported in column 4 of the Table~\ref{tab11}
are the observed values.

A comparison of our estimates of $\Delta m_{15}(B)$ and colour excess
with those reported by \cite{phil99} for the objects in common shows
that, except for a few cases, the values agree within the errors. The
differences are due to slightly different choices for the epoch and
magnitude at maximum.

\begin{figure*}
\includegraphics[width=58mm]{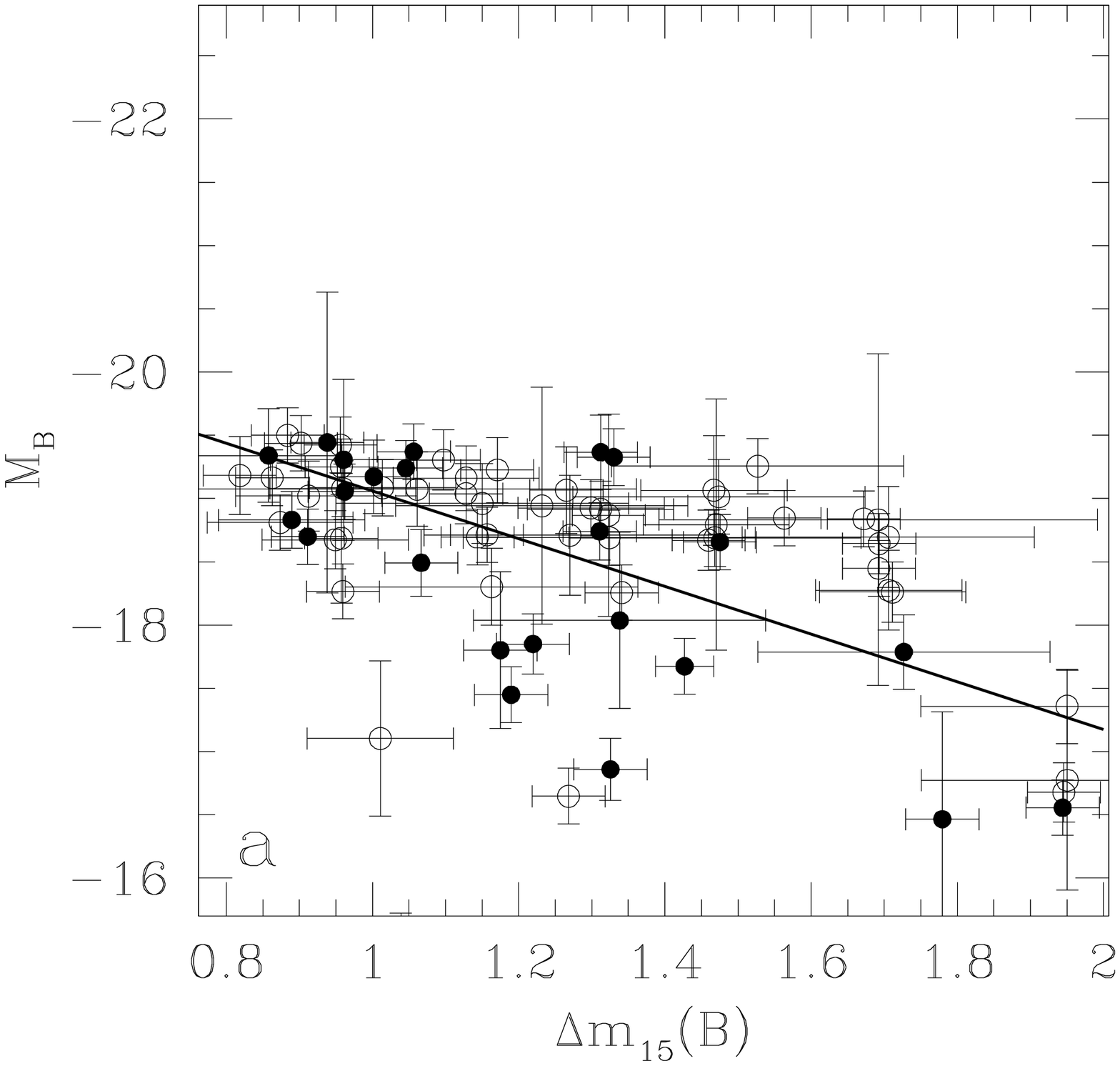}
\includegraphics[width=58mm]{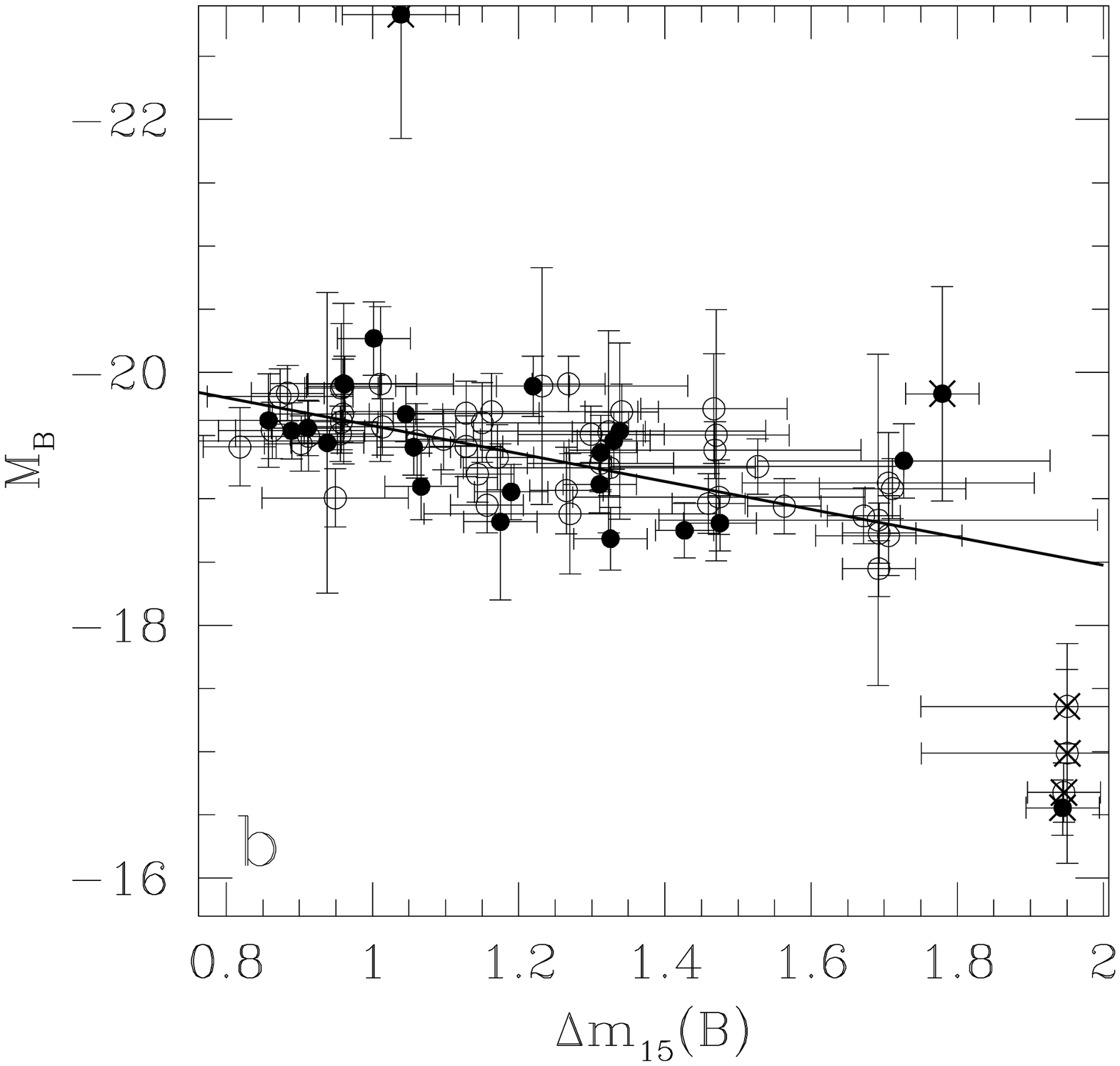}
\includegraphics[width=58mm]{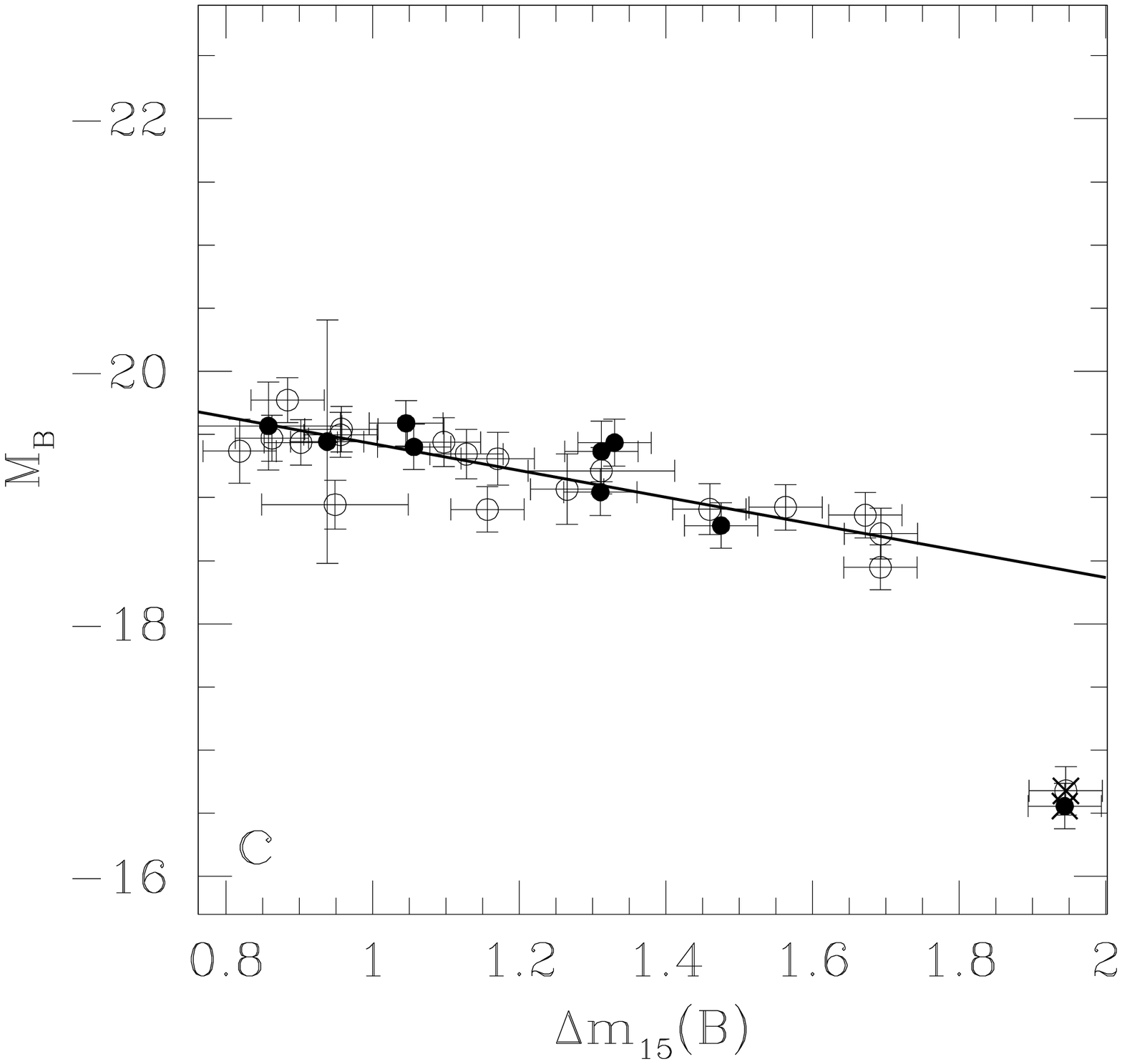}
\caption{$M_B$ versus $\Delta m_{15}(B)$ relation; filled circles
 indicate objects whose distances are given by Tully's catalogue, open
 symbols are objects whose distances are calculated from their recession
 velocity.  The linear fit is weighted in both axes
 (\citealt{numericalrecipes}).  
{\textbf{a}}: Only galactic reddening correction applied. Number of objects
 used for the linear fit n=73, dispersion $\sigma=0.83$.
{\textbf{b}}: Both Galactic and host galaxy reddening corrections
applied.  The outliers, marked with a cross, are (from left to right,
from top to bottom): SN 1996ai, which is characterized by high and not
well known reddening; SN 1986G, another highly reddened event; SN
1992K, 1999da, 1998de, 1991bg, which are peculiar sub-luminous
events. The latters seem to form a separate class and do not fit the
linear relation defined by all others. n=67, $\sigma=0.31$.
% obtained with fitexy.f 
{\textbf{c}}: as the previous case but  selecting only SNe with $E(B-V)<0.1$
and small errors ($<0.2$)  in $\Delta m_{15}(B)$. n=26, $\sigma=0.20$, $R_B=3.5$.}
\label{MBdm15ps}
\end{figure*}

\subsection{Absolute magnitude vs. $\Delta m_{15}(B)$}\label{mbdm15}

In order to use SNe Ia as distance indicators, their absolute
magnitude $M_B$ needs to be calibrated. This requires that at least
for a sub-sample of objects, the distance modulus of the host galaxies
is measured by means of some independent distance indicators. As we
will argue in the next section, this is best done with Cepheids.

If SNe Ia were true standard candles, the calibrated absolute
magnitude could be directly used to measure the distance of all other
SN host galaxies.  As we mentioned before, this is not the case  but,
thanks to the relation between absolute magnitude at maximum and
luminosity decline \citep{phil93}, the role of SNe Ia as distance
indicators can be recovered.  Therefore is this relation which needs
to be calibrated.

Unfortunately this cannot be done using the Cepheid calibrated  SNe Ia alone
for two reasons:

\begin{enumerate}
  \item the number of SN Ia host galaxies with Cepheid measured
  distances is too small (only 9 objects, as shown in
  Table~\ref{tabellacef});
 
  \item as shown in Fig.~\ref{dm15tipogalassia}, SNe Ia with different
  absolute magnitudes are not uniformly distributed versus galaxy
  types. In particular faint, rapidly declining SNe Ia occur
  preferentially in early type galaxies (see also \citealt{filippo89},
  \citealt{hamuy96c}, \citealt{cappellaro2001}).  As discussed by
  \citet{ivanov2000} this effect may be due to the age of the
  progenitors (the older the progenitor system, the fainter the
  corresponding SN), but also to metallicity (lower metallicity
  systems produce fainter SNe, \citealt{umeda99}).

  Since Cepheids are found in spirals only, it is not possible to
  properly sample the $M_B$ -- $\Delta m_{15}(B)$ relation by means of
  Cepheid calibrated SNe Ia.
\end{enumerate}

To overcome these difficulties we adopted a two-step procedure. We
first determine the slope of the relation using some widely available,
though less accurate, distance indicators (Fig.~\ref{MBdm15ps});
then, since the linear fit depends on the assumed value of $H_0$ ($75$
\,km\,s$^{-1}$Mpc$^{-1}$), we determine the zero point by minimizing
the deviation of the Cepheids calibrated SNe Ia on this fixed slope
(see next section).

\begin{table*}
\begin{center}
\caption{
Parameters of the $M_B$=a($\Delta m_{15}(B)-1.1)+b$  relation.
From top to bottom:
values obtained for case {\textbf b}, {\textbf c} (as in \S \ref{mbdm15}).
From left to right: number  of objects used for the correlation; $R_B$ adopted;
slope (error); zero point  derived  from three different assumptions on the metallicity  
PL relation (error);  dispersion.
}\label{tab1}
\begin{tabular}{crcccccc}
\hline
\hline
 n & $R_B$ &  a & \multicolumn{4}{c}{b}& $\sigma$ \\
\cline{4-7}
           &   &    &    KP & ${\Delta Y}/{\Delta Z}=2.5$ &
${\Delta Y}/{\Delta Z}=3.5$ &&  \\
\hline
67 & 4.315 & 1.102 (0.147) &  -19.613 &-19.523& -19.582&(0.037)   & 0.31     \\
67 & 3.5   & 1.092 (0.124) &  -19.460 &-19.399& -19.472&(0.031)   & 0.28     \\
%26 & 4.315 & 1.082 (0.183) &  -19.615 &-19.525& -19.584&(0.053)   & 0.20     \\
26 & 3.5   & 1.061 (0.154) &  -19.455 &-19.403& -19.476&(0.044)   & 0.20     \\
\hline
\hline
\end{tabular}
\end{center}
\end{table*}

For the first step, absolute magnitudes have been calculated using
distances obtained from two different sources.  Distance moduli for 24
host galaxies were retrieved from the {\it ``Nearby Galaxies
Catalog''} (\citealt{tully88}). For the galaxies not listed there and
with recession velocity larger than 3000 km\,s$^{-1}$ (49 objects), we
have determined the distance from the recession velocity, adopting
$H_0=75$ \,km\,s$^{-1}$Mpc$^{-1}$ to be consistent with the Tully's
catalogue scale.  In order to avoid contamination from peculiar
motions, we rejected 5 of the 78 SNe of the sample which are not
listed in Tully's catalogue and have low recession velocity.
Recession velocities in the reference frame of the 3K background
microwave radiation were retrieved from the 
NASA/IPAC Extragalactic Database\footnote{http://nedwww.ipac.caltech.edu/}.

\begin{figure}
\begin{center}
\includegraphics[angle=0,width=80mm]{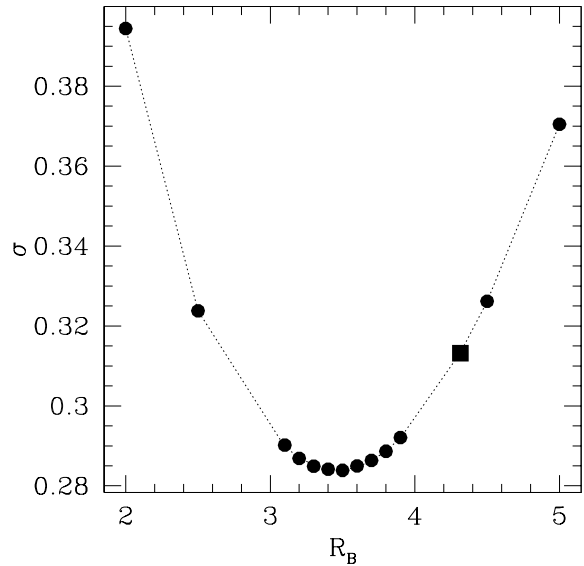}
\caption{Dispersion of the $M_B$ -- $\Delta m_{15}$ relation for
different values of $R_B$.  The  sample is the same as in figure
\ref{MBdm15ps}~{\textbf{b}}.  The filled square corresponds to
$R_B=4.315$.  The minimum corresponds to $R_B=3.5$.}
\label{RB_sigmaminima}
\end{center}
\end{figure}

\begin{figure}
\begin{center}
\includegraphics[angle=0,width=80mm]{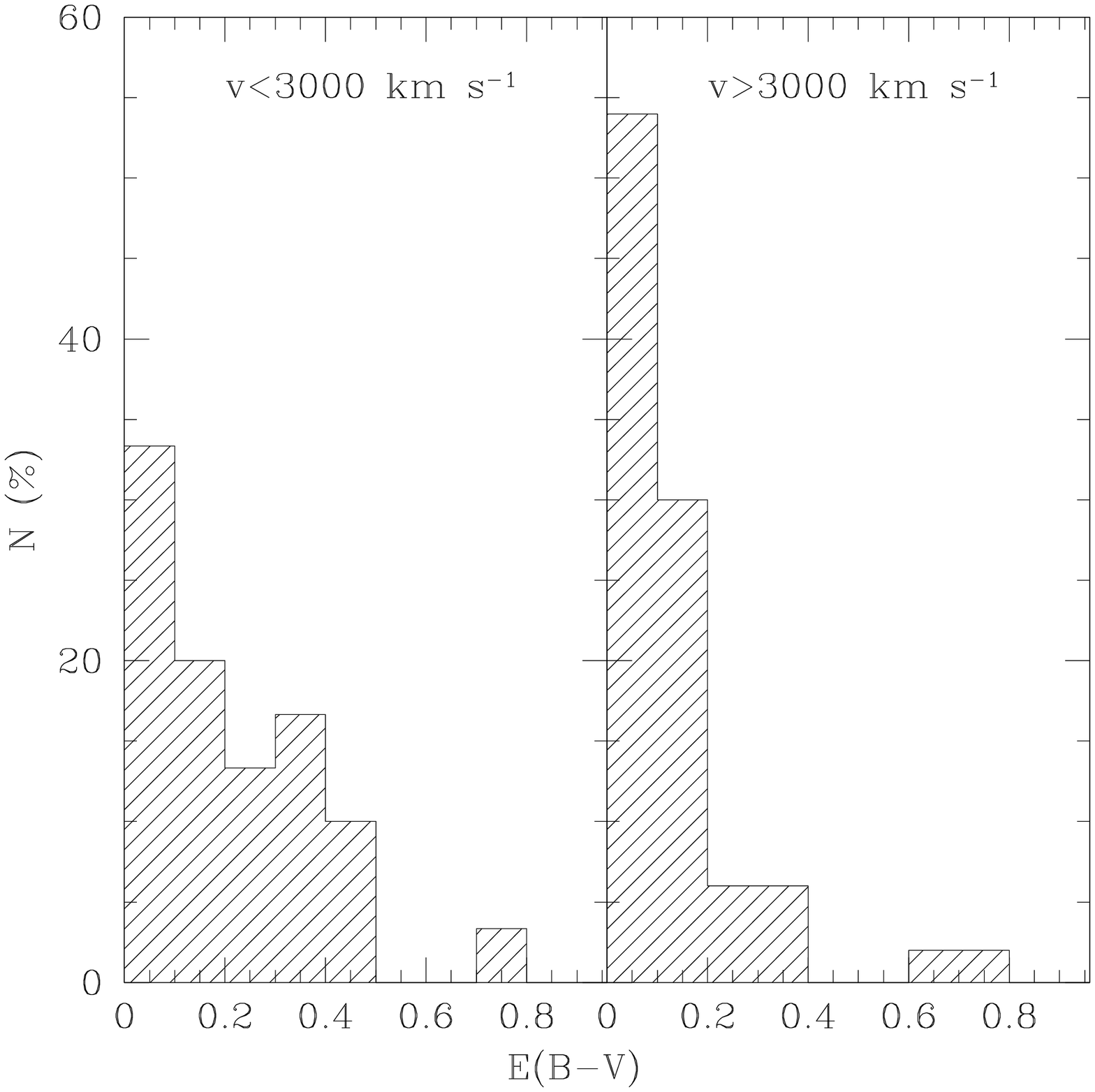}
\caption{Distribution of the host galaxy extinction $E(B-V)_{host}$. 
Left: SNe with recession velocity smaller 
than 3000 km\,s$^{-1}$; Right:  SNe with recession velocity bigger than 3000 
km\,s$^{-1}$. }
\label{ebv_z}
\end{center}
\end{figure}

To show the importance of the host galaxy extinction correction, in
Fig.~\ref{MBdm15ps}{\textbf a} we have plotted the data of 73
SNe, corrected for Galactic extinction only. As it can be seen, the
the points show a large dispersion and the correlation between $M_B$
and $\Delta m_{15}(B)$ remains hidden.

The scatter appears greatly reduced in Fig.~\ref{MBdm15ps}{\textbf b}
where the SN magnitudes have been corrected for extinction in the host
galaxy.  The latter has been derived from the colour excess estimated
in \S \ref{Reddening} and assuming a standard reddening law, that is
$A_B=4.315 \times E(B-V)$ (\citealt{schlegel}, see also \citealt{cardelli89}).  
The only evident
outliers are now four fast declining SNe (the sub-luminous SNe 1992K,
1999da, 1998de, 1991bg), SN 1996ai, whose colour excess is very high
but not well defined \citep{phil99}, and 1986G, another highly
reddened SN.

Actually, we found that the dispersion in the $M_B$ -- $\Delta m_{15}(B)$
relation can be further reduced assuming a different value for $R_B$.
As shown in Fig.~\ref{RB_sigmaminima}, a minimum dispersion is
obtained for $R_B=3.5$, somewhat smaller than the canonical values
$R_B=4.315$. 

A qualitatively similar result was obtained, again using SN Ia, by 
\cite{capaccioli90}, who favoured $R_B=1.7$. However,  
previous measurements of low values for $R_B$ from SNe were plagued by an 
inadequate understanding of the relation between intrinsic color
and luminosity at maximum \citep{riess96b}. 
Other measurements of absorption in nearby galaxies provided values for 
$R_B$ ranging from the canonical value ($R_B\sim 4.3$) to much smaller 
values ($R_B\sim 3.6$: 
\citealt{bouchet1985}, \citealt{brosch88}, \citealt{brosch91}; $R_B\sim 
2.4$:
\citealt{rifatto90}). However a lower estimate for the host galaxy 
reddening correction $R_B$ has been suggested also by  more recent studies  
($R_B\sim 3.5$ \citealt{phil99,knop2003}).
\\
Actually, even in our own Galaxy  the ratio of the total to selective 
absorption has been found to vary significantly from place to place (e.g. 
$R_V$ varies from 2.6 to 5.5 in the measurements of  \citealt{clayton88}). 
Therefore, in principle, it would be interesting  to study the dependence 
of $R_B$ on other observables, such as the distance from the host galaxy 
centre or the integrated galaxy properties. This would require the 
possibility to derive
estimate of $R_B$ for individual objects which, unfortunately, cannot be 
done with the present data.
\\ 
We note that,
from a practical point of view, reducing $R_B$ has the same effect of
the adoption of a relatively narrow a-priori distribution of the
reddening for a Bayesan estimate of the individual SN extinction
(cf. \S\ref{Reddening}). 
In \S\ref{hubble} we will check what is the
effect on the Hubble diagram of this different choice for $R_B$.

The issue is important (see also \citealt{fischer03}) because it turns
out that, on average, the more distant SNe of our sample are less
affected by the extinction than nearby ones. This is shown in
Fig.~\ref{ebv_z}: it turns out that for SNe in galaxies with recession
velocity smaller than 3000 $km\,s^{-1}$ the average colour excess is $\sim 0.27$,
whereas the same number for more distant galaxies is $\sim 0.13$.

This is likely due to selection effects, since distant obscured
objects are more difficult to detect and SNe are likely to be
discovered far from their host nucleus in distant galaxies. 

As a final test we plot the $M_B$ -- $\Delta m_{15}(B)$ for the subsample
of 26 SNe with $E(B-V)<0.1$ and well measured $\Delta m_{15}$ ($\Delta
m_{15}$ uncertainties $<0.2$) (see also \citealt{phil99},
\citealt{freedman2001}). This has two advantages, it gives a more
homogeneous sample and the uncertainties related to the host galaxy
extinction correction are reduced. The result, shown in
Fig.\ref{MBdm15ps}{\textbf c}, is a very tight relation with
dispersion $\sigma=0.20$.

\begin{table*}
\begin{center}
\caption{Cepheid calibrated SNe.  (1) SN name; (2) host galaxy; (3)
  apparent B magnitude at maximum; (4) $B-V$ colour at maximum; (5)
  $\Delta m_{15}(B)_{obs}$ (not corrected for colour excess); (6)
  $E(B-V)_{gal}$ from \citealt{schlegel}; (7) $E(B-V)_{host}$: weighted
  mean between $E(B-V)_{max}$ and $E(B-V)_{tail}$; (8) $\mu_0$(KP),
  (9) $\mu_0(\Delta Y/\Delta Z)=2.5$, (10) $\mu_0(\Delta Y/\Delta
  Z)=3.5$ are the distance moduli given by the HST Key Project
  \citep{freedman2001} and the distance moduli obtained for two
  different metallicity (see \S \ref{cefeidisection}).  }
\setlength\tabcolsep{2pt}
\label{tabellacef}
\begin{tabular}{lccccccccc}
\hline
\hline
 SN   & galaxy   &  m(B)$_{max}$  & $B-V$ & $\Delta m_{15}(B)_{obs}$ & $E(B-V)_{gal}$ & $E(B-V)_{host}$ & $\mu_0$(KP) &  $\mu_0(\frac{\Delta Y}{\Delta Z}=2.5)$ & $\mu_0(\frac{\Delta Y}{\Delta Z}=3.5)$\\
 (1) & (2) & (3) & (4) & (5) & (6) & (7) & (8) & (9) & (10) \\
\hline
1990N  & NGC4639  & 12.75(0.04)  &  0.04(0.05) &1.05(0.05)  &    0.026(0.003)  &  0.14(0.06)  &   31.61(0.08)   &   31.47(0.26)    & 31.55(0.31)\\
1981B  & NGC4536  & 12.04(0.04)  &  0.06(0.05) &1.13(0.05)  &    0.018(0.002)  &  0.11(0.05)  &   30.80(0.04)   &   30.66(0.20)    & 30.71(0.23)\\
1989B  & NGC3627  & 12.38(0.12)  &  0.38(0.05) &1.28(0.05)  &    0.032(0.003)  &  0.42(0.05)  &   29.86(0.08)   &   29.88(0.42)    & 30.11(0.54)\\
1998bu & NGC3368  & 12.20(0.04)  &  0.34(0.05) &1.15(0.05)  &    0.025(0.003)  &  0.37(0.05)  &   29.97(0.06)   &   29.94(0.50)    & 30.12(0.63)\\
1972E  & NGC5253  &  8.40(0.04)  & -0.06(0.10) &1.05(0.05)  &    0.056(0.006)  &  0.01(0.05)  &   27.56(0.14)   &   27.65(0.15)    & 27.65(0.14)\\
1937C  & IC4182   &  8.94(0.30)  &  0.00(0.20)  &0.85(0.20) &    0.014(0.001)  &  0.06(0.05)  &   28.28(0.06)   &   28.31(0.08)    & 28.31(0.08)\\
1960F  & NGC4496  & 11.65(0.20)  &  ...        &1.10(0.20)  &    0.025(0.005)  &  ...         &   30.81(0.03)   &   30.70(0.18)    & 30.72(0.20)\\
1974G  & NGC4414  & 12.34(0.04)  &  0.22(0.20) &1.40(0.04)  &    0.019(0.005)  &  0.25(0.05)  &   31.10(0.05)   &   31.07(0.38)    & 31.25(0.48)\\
1991T  & NGC4527  & 11.69(0.04)  &  0.14(0.05) &0.94(0.05)  &    0.022(0.002)  &  0.20(0.05)  &   30.74(0.12)   &   30.60(0.57)    & 30.65(0.66)\\
\hline
\hline
\end{tabular}
\end{center}
\end{table*}

The comparison of  our results, reported in Table~\ref{tab1}, with those obtained from
26 SNe of the Cal{\'a}n/Tololo survey \citep{hamuy96c}, indicates that we
find a slightly steeper slope, in the range $1.061\pm0.154$ to $1.102 \pm
0.147$ depending on the sample vs $0.784 \pm 0.182$ of \cite{hamuy96c}. It is
interesting to note that the difference disappear if, as in
\cite{hamuy96c}, we neglect the host galaxy extinction correction
($a=0.855 \pm 0.182$). %($a=0.855 \pm 0.182$) per RB=4.315, 0.878 +/- 0.153 per RB=3.5 both for the low extinction sample

The next step is determine the zero point of the $M_B$ -- $\Delta
m_{15}(B)$ relation by means of Cepheids calibrated SNe.

\subsection{Effect of the metal content on the Cepheids-distance scale}\label{cefeidisection}

\begin{figure}
\begin{center}
\includegraphics[width=84mm]{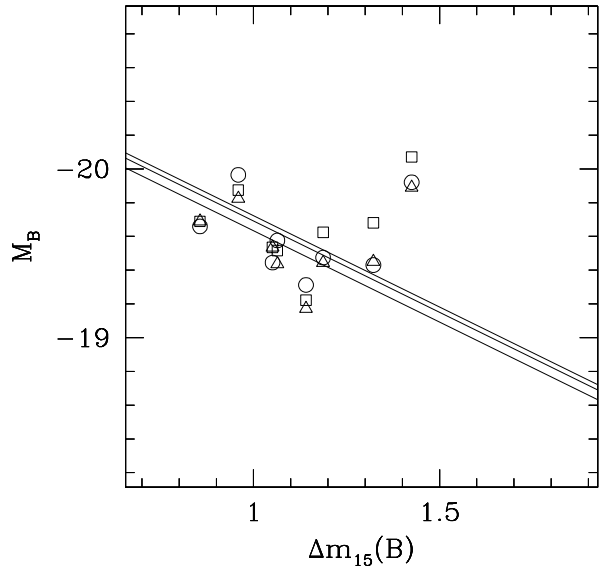}
\caption{$M_B$ versus $\Delta m_{15}(B)$ for 8 of the 9 SNe calibrated
by Cepheids (SN 1960F has been excluded because colour information
required to estimate the host galaxy extinction is not available).
Circles: Distance modulus $\mu$ from Freedman et al. 2001; triangles:
$\mu$ corrected assuming $\Delta Y/\Delta Z=2.5$; square: $\mu$
corrected assuming $\Delta Y/\Delta Z=3.5$.  Solid lines are the best
fits obtained with a fixed slope minimizing the deviation of the three
sets of Cepheids calibrated SNe. The slope (1.082) is the same as in
Fig.~\ref{MBdm15ps}{\textbf c}.  }
\label{cefeidi}
\end{center}
\end{figure}

Classical Cepheids are fundamental primary distance indicators thanks
to their characteristic Period-Luminosity (PL) and
Period-Luminosity-Colour (PLC) relations. Such relations, once
calibrated, allow one to evaluate distances of the order of 3 Mpc from
ground-based observations and about 30 Mpc from space observations
(HST, \citealt{freedman2001} and references therein). Moreover, the
Cepheid distance scale constrains the evaluation of the Hubble
constant $H_0$ through the calibration of secondary distance
indicators.  This implies that any systematic error on the Cepheids as
standard candles may affect the inferred extragalactic distance scale
and, in turn, the value of $H_0$.

In the recent years, one of the most debated issues concerning the
extragalactic distance scale is the possible dependence of Cepheid
properties on the chemical composition of the host stellar
population. Indeed, Cepheid PL relations are traditionally considered
to be ``universal'' \citep{iben84,freedman90}, with the slope derived
from Cepheids in the Large Magellanic Cloud (LMC),
\citep{madore91,udalski2000} and the zero point corresponding to some
assumption on the LMC distance modulus or on Galactic Cepheids with
independent distance estimates.  Empirical tests of the metallicity
effect on the Cepheid distance scale suggest that either the effect is
very small (\citealt{freedman90}) or it goes in the direction of
predicting brighter Cepheids for higher metallicities
(\citealt{kennicutt98}).

On the basis of the latter test, the HST Key Project (KP) distance
moduli are corrected for the metallicity effect by adopting the
correction $\Delta{\mu_0}/\Delta{[O/H]}=0.2\pm0.2$ (Freedman et
al. 2001).

\begin{figure*}
\begin{center}
\includegraphics[width=58mm]{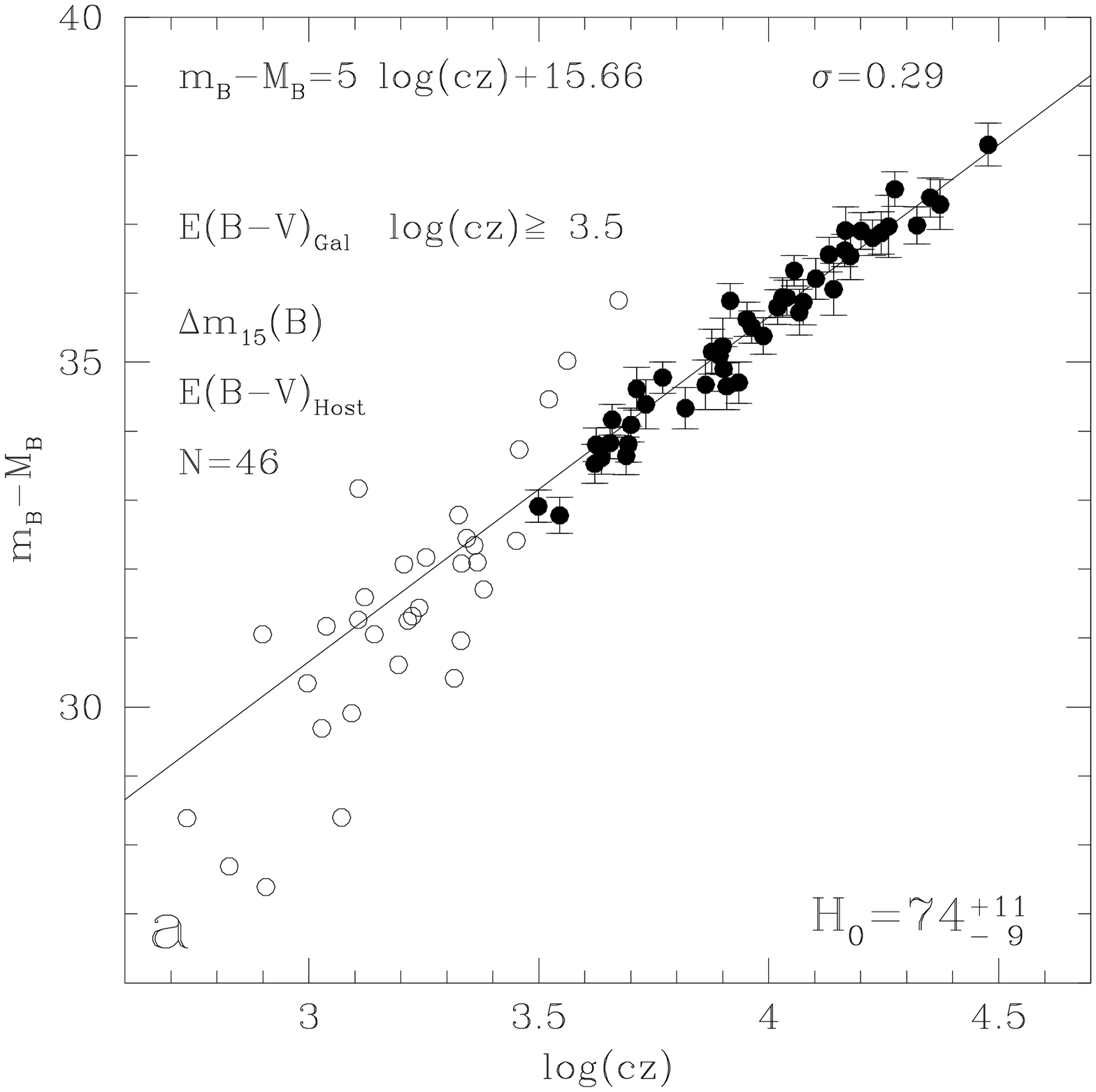}%{hubble11b.eps}
\includegraphics[width=58mm]{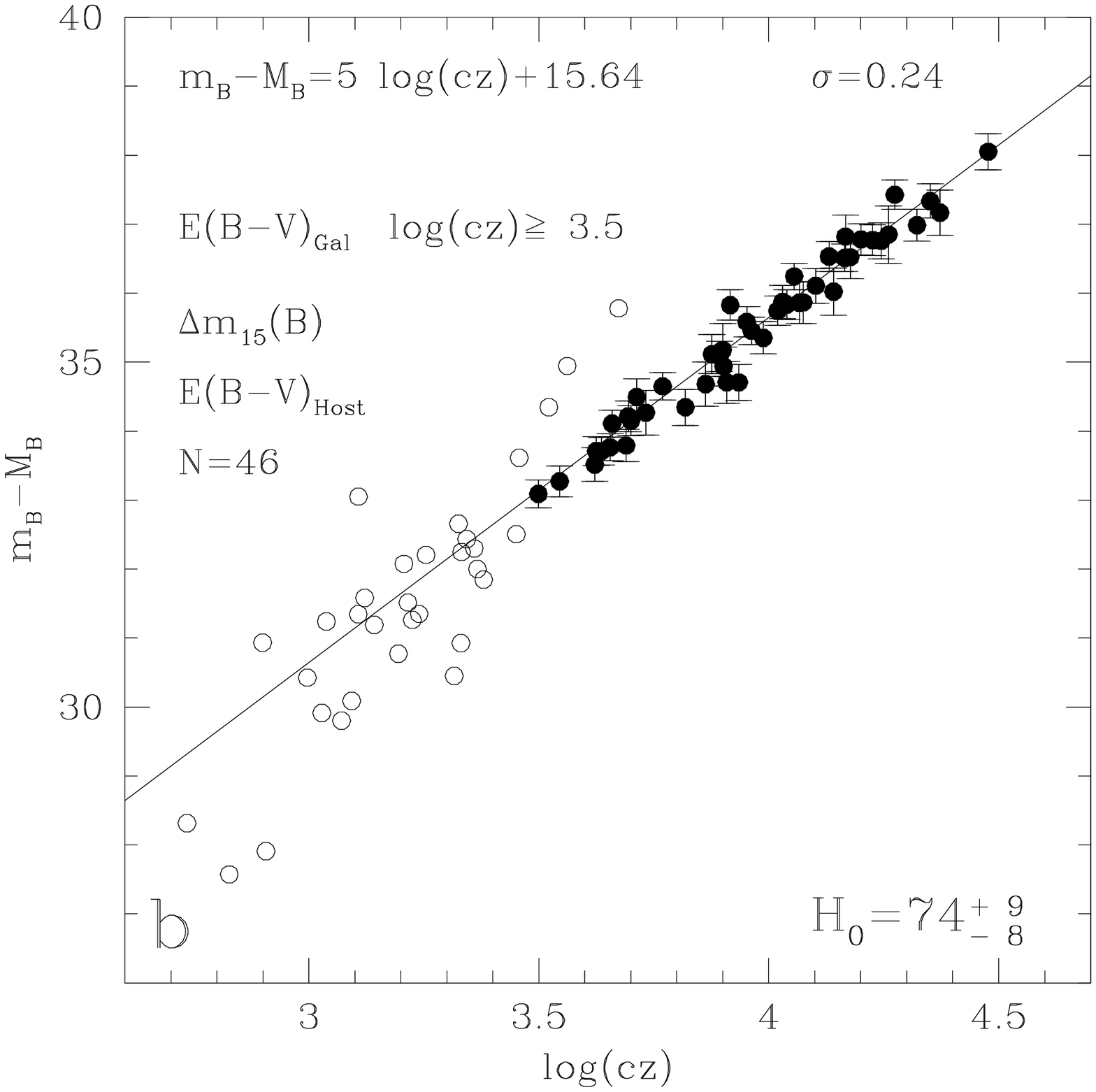}%{hubble3.5.eps}
\includegraphics[width=58mm]{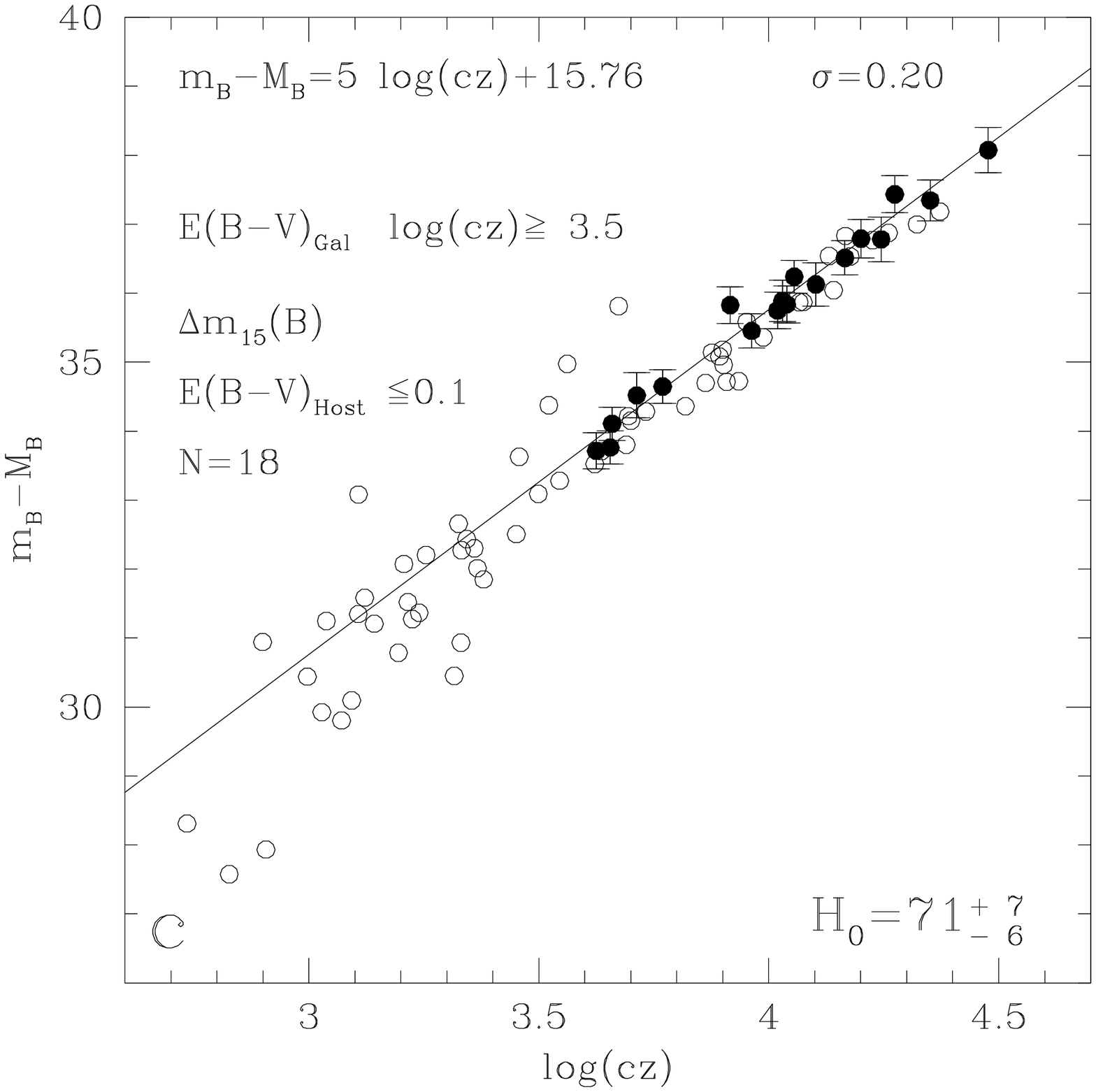}%{hubble3.5_2.eps}
\caption{ {\textbf{a}}: Hubble diagram of the whole sample assuming
$R_B=4.315$ in the estimate of the host galaxy absorption.
{\textbf{b}}: as the previous plot but with $R_B=3.5$. {\textbf{c}}:
Hubble diagram for the good photometry, low extinction sample ($R_B=3.5$). 
In all plots, we have used for the fit only those SN with recession
 velocity $>$ 3000 km s$^{-1}$ and we have adopted the Cepheids P-L correction for
 $\Delta{Y}/\Delta{Z}$=2.5.  The fast declining SNe 1991bg, 1992K,
 1999da and 1998de have been excluded from the fit. 
Points which do not contribute to the fit are
indicated as open circles.}
\label{hubbledia}
\end{center}
\end{figure*}

From the theoretical point of view, nonlinear convective pulsation
models (\citealt{bono99a}), which are able to predict all the relevant
pulsation observables, suggest that in the range $0.004 < Z < 0.02$
metal poor Cepheids are brighter than the metal-rich ones at fixed
pulsation period \citep{bono99b,caputo00a}.  As a consequence, at
least in this range of Z, the adoption of ``universal'' LMC-based PL
relations provides distances that are systematically overestimated
(underestimated) for galaxies more metal-rich (metal-poor) than the
LMC \citep{caputo00b}.  On the other hand, when the pulsation analysis
is extended to metallicities higher than the solar value (up to
Z=0.04) in order to cover the whole range of metal abundances of the
HST Key Project galaxies \citep{ferrarese00}, and the simultaneous
effect of the increase in the helium abundance is taken into account,
the predicted metallicity correction changes sign around the solar
value and depends significantly on the adopted helium to metallicity
enrichment ratio $\Delta{Y}/\Delta{Z}$ (see \citealt{fiorentino02} for
details).  In particular \citet{fiorentino02} find that for
$\Delta{Y}/\Delta{Z}$=3.5, the theoretical metallicity correction is
in agreement with the result of the empirical test by
\citet{kennicutt98}.

In order to calibrate the $M_B$-$\Delta m_{15}(B)$ relation, we have
considered three different estimates for the Cepheid distance to the
SN host galaxies: $\mu_0$(KP), computed without metallicity correction, and 
$\mu_0({\Delta Y}/{\Delta Z}=2.5)$, $\mu_0({\Delta Y}/{\Delta Z}=3.5)$
 with the metallicity corrections  for two different
choices of the adopted helium to metallicity enrichment ratio
$\Delta{Y}/\Delta{Z}$ 
(Table~\ref{tabellacef}).  The zero point of the $M_B$ -- $\Delta
m_{15}(B)$ relation was determined by minimizing the deviation of the
linear fit for the Cepheids calibrated SN, maintaining a fixed slope
(Fig.~\ref{cefeidi}).  The inferred parameters of the correlation are
reported in Table~\ref{tab1}. The zero point obtained with different
assumptions in the calibration lies in the range $-19.613<b<-19.399$,
to be compared with $b=-19.258 \pm 0.048$ in \citealt{hamuy96c}.

\section{The Hubble Constant}\label{hubble}

After completion of the calibration path, we can now use SNe~Ia to
measure the Hubble constant.

In Fig.~\ref{hubbledia} we show three examples aimed to show how
different corrections and sample selection affect the scatter of the
Hubble diagram. For this purpose, we adopted the PL relation corrected
for $\Delta{Y}/\Delta{Z}$=2.5 (see Table~\ref{tab1}) and we excluded
from the fit all SNe Ia with recession velocity $v < 3000$
km~s$^{-1}$.  The latter is intended to avoid contamination by
peculiar motions, which, on average, amount to $\sim$200--300\,km\,s$^{-1}$ 
(\citealt{tonry2000}, \citealt{giovannelli99}).
We have also excluded SNe with $\Delta m_{15}(B)> 1.8$ because they do
not fit the linear relation $M_B-\Delta m_{15}(B)$, as seen in
Fig.~\ref{MBdm15ps}{\textbf{b}}.

In panel $a$ of Fig.~\ref{hubbledia} is the Hubble diagram obtained
using the complete sample and a standard reddening law, in panel $b$
is the plot obtained adopting $R_B=3.5$ for the ratio of total to
selective extinction (cfr. \S\ref{mbdm15}) and finally, in panel $c$ is
shown the Hubble diagram for the sample of SN~Ia with the best
photometric coverage $\Delta m_{15}(B) < 0.2$ and low reddening
$E(B-V)_{host}>0.1$ ($R_B=3.5$).

The fact that the dispersion of the points in panel $b$
($\sigma=0.24$) is significantly reduced compared with panel $a$
($\sigma=0.29$) is consistent with the finding of
\S\ref{mbdm15}. Again, it appears that the host galaxy absorption
correction deduced from the measured colour excess and adopting a
standard reddening law is an overestimate of the true absorption. On
the other side, this turns out to have a small  effect on the value of
the Hubble constant itself ($H_0=74^{+9}_{-8}$ in panel $b$ compared
with $H_0=74^{+11}_{-9}$ in panel $a$) provided that in all steps the
calibration of the SN magnitudes is done consistently (see also Table~\ref{tab2}).

The dispersion of the Hubble diagram is further reduced by selecting
the subsample of SN Ia with good photometry and low reddening
SNe ($\sigma=0.20$). This time the Hubble constant is slightly reduced
$H_0=71^{+7}_{-6}$.  Although this is our preferred result and that
with the smallest statistical uncertainty, we stress that the
selection criteria were not applied to the Cepheid calibrated SN of
Table~\ref{tabellacef} which otherwise would be reduced to only two
cases. In principle, this asymmetry may introduce some bias.

The different results are summarized in Table~\ref{tab2} where in
particular it can be seen the sensitivity of the estimate of the
Hubble constant on the different corrections of the PL relation. As it
can be seen the effect is, in all cases, quite small.

$H_0$ ranges between 68 and 74\,km\, s$^{-1}$ Mpc$^{-1}$, with
uncertainties of the order of $\sim 7$ \,km\, s$^{-1}$ Mpc$^{-1}$
($\sim 10\%$)\footnote{The fact that the distance scale turned out to
be very close to that of the Nearby Galaxies Catalogue by
\citet{tully88} which was used to derive the slope of the
$M_B$-$\Delta m_{15}(B)$ relation assure on the negligible impact of
possible second order effects.}

These values are in agreement with the result of the Hubble Space
Telescope Key Project (\citealt{freedman2001}) which gives $H_0=72\pm
8$ km \,s$^{-1}$ Mpc$^{-1}$.  They are also in agreement with the
recent results of {\it WMAP}, namely $H_0=72 \pm 5$ km \,s$^{-1}$
Mpc$^{-1}$ \citep{spergel2003}, obtained for a spatially flat
cosmological model dominated by cold dark matter and non-zero
cosmological constant ($\Lambda$CDM model). Note that the same {\it
WMAP} data give smaller value for $H_0$ if fitted by different
cosmological models.

\section{Conclusions}\label{conclusion}

Using a large sample of SNe Ia light curves, both from the literature
or unpublished data, we have revisited
the different steps required to calibrate the SN absolute
magnitude and derive the Hubble constant. 

In particular, we have re-calibrated the relation between the SN absolute
magnitudes and their rate of decline soon after maximum. The latter can
be measured without distinction by the $\Delta m_{15}(B)$ or by the
``stretch factor''

SN~Ia with Cepheids calibrated distances are crucial because they are
used to calibrate the zero point of the $M_B$ vs. $\Delta m_{15}(B)$
relation.  We have considered the effect of a new metallicity
correction of the Cepheids PL relation, how this reflects on the 
calibration of SN~Ia and finally on the estimate of the Hubble constant.
Fortunately, the latter turned out to be not very important.

Instead we found that a major uncertainty is related to the correction
of the extinction in the host galaxy.  In particular, we found that
the scatter of the $M_B$ -- $\Delta m_{15}$ relation and of the Hubble
diagram itself can be reduced assuming a (slightly) smaller value for
the ratio of the total to selective absorption, $R_B$. Indeed, the
scatter is minimum for $R_B=3.5$.  The difference with the standard
value, $R_B=4.315$, is well within the variance of the measurements of
this parameter even within our own Galaxy. Therefore it does not
require different properties for the dust in other galaxies. Still
this is an important issue for the use of SN as distance indicators.

The Hubble constant, obtained with different assumptions on the
metallicity dependence of the Cepheid PL relation, different cuts in
the sample and varying the value of $R_B$, lies in the range 68--74
\,km\,s$^{-1}$Mpc$^{-1}$, with uncertainties of the order of 10\%.

\begin{table}
\begin{center}
\caption{Values of $H_0$ (in km s$^{-1}$ Mpc$^{-1}$) for different
assumptions on the Cepheid calibration and $R_B$.  $a$ and $b$  are the values  for 
the complete sample, whereas in $c$   are the values for the 
good photometry, low reddening sample.}
\label{tab2}
\begin{tabular}{lcccc}
\hline
                &     \textbf{a}  & \textbf{b} &   \textbf{c}      \\
         N(SN)  &       46        &      46    &         18         \\
        $R_B$   &      4.315      &      3.5   &        3.5         \\
\hline

$\mu_0$(KP)     
                &  $71^{+10}_{-9}$&  $72^{+8}_{-7}$ &  $69^{+7}_{-6}$\\
\\  
$\mu_0(\frac{\Delta Y}{\Delta Z}=2.5)$ 
                &  $74^{+11}_{-9}$&  $74^{+9}_{-8}$ &  $71^{+7}_{-6}$\\
\\
$\mu_0(\frac{\Delta Y}{\Delta Z}=3.5)$ 
                &  $72^{+10}_{-9}$&  $72^{+8}_{-7}$ &  $68^{+6}_{-6}$\\
%                &     \textbf{a}  & \textbf{b} &  \textbf{c}  &  \textbf{d}      \\
%         N(SN)  &       46        &      46    &       18     &       18         \\
%        $R_B$   &      4.315      &      3.5   &    4.315     &      3.5         \\
%\hline%
%
%$\mu_0$(KP)     
%                &  $71^{+10}_{-9}$&  $72^{+8}_{-7}$ & $65^{+6}_{-6}$ & $69^{+7}_{-6}$\\
%\\  
%$\mu_0(\frac{\Delta Y}{\Delta Z}=2.5)$ 
%                &  $74^{+11}_{-9}$&  $74^{+9}_{-8}$ & $68^{+6}_{-6}$ & $71^{+7}_{-6}$\\
%\\
%$\mu_0(\frac{\Delta Y}{\Delta Z}=3.5)$ 
%                &  $72^{+10}_{-9}$&  $72^{+8}_{-7}$ & $66^{+6}_{-6}$ & $68^{+6}_{-6}$\\
\hline
\hline
\end{tabular}
\end{center}
\end{table}

\section{Acknowledgments} 
Unpublished data are based  on observations collected at ESO - La Silla (Chile), Asiago (Italy), TNG 
(La Palma, Canary Islands, Spain) and HST WFPC2 Archive.
We acknowledge support from the Italian Ministry For Instruction, 
University and Research (MIUR) through grant Cofin 2001021149-002
and from the European Research Training Network ``The Physics of Type Ia Supernova'' 
through grant RTN 2-2001-00037. 
This research has made use of the NASA/IPAC Extragalactic  
Database 
(NED) which is operated by the Jet Propulsion Laboratory, California Institute of
 Technology, under contract with the National Aeronautics and Space Administration. 

%\bibitem[\protect\citeauthoryear{}{}]{}

\clearpage
\newpage 
\appendix 
\onecolumn
\section{ESO - Asiago photometric data}\label{fotometriapadova}
Table~\ref{esoasiagoarchive} reports the unpublished photometry based on observations collected
mainly at ESO - La Silla (Chile) and Asiago (Italy), plus some data from TNG (La Palma, Canary 
Islands, Spain) and from the HST WFPC2 Archive.
\\
Data reduction has been performed using IRAF\footnote{Image Reduction and Analysis Facility, 
http://iraf.noao.edu/} standard  recipes for trimming, bias and overscan correction and
flat fielding.
\\
SN magnitudes have been measured mainly using a  Point Spread Function (PSF)  fitting technique,
which ensures a  good removal of bad pixels and cosmic rays \citep{turatto93}.
\\
Photometric calibration was obtained from observation of Landolt standard stars \citep{landolt92} 
in photometric nights. This was also used to establish a local sequence which was used to calibrate
the other observations.
\\
In  brackets are   indicative estimates of the errors, 
obtained from the values measured on several reference nights through artificial stars experiments,
i.e. adding a fake star, as bright as the SN, in several different positions around the SN location, 
and measuring the spread of the resulting magnitudes  with the same procedure used for the SN. 
\\
The codes for the telescope/instrument used are as follows:
\\
1= Danish 1.54m + DFOSC, 2= MPG/ESO 2.2m + EFOSC2, 3= ESO 3.6m + EFOSC1, 4= NTT + SUSI, 5= Dutch 0.91m,
6= NTT + EMMI, 7= Asiago 1.82m + AFOSC, 8= ESO 3.6m + EFOSC2, 9= TNG + OIG, 10= ESO 1.0m, 
11=  HST WFPC2 Archive.
\floatplacement{table}{h}
\begin{table*}
\begin{center}
\caption{}
\label{esoasiagoarchive}
\begin{tabular}{l l c c c c c l}
\hline
\hline
\noalign{\smallskip}
Date & Julian Day & $U$ & $B$ & $V$ & $R$ & $I$ & telescope  \\
     & {\footnotesize +2,400,000} &     &     &     &     &     &             \\
\noalign{\smallskip}
\hline
\noalign{\smallskip}
\multicolumn{8}{c}{SN 1991S }\\
\noalign{\smallskip}
 13/04/91 &    48359.67 &     -   &      -    &  18.15(0.03) &      -    &        -    &    1 \\
 18/04/91 &    48364.51 &     -   & 18.94(0.02) &  18.42(0.03) & 18.31(0.02) &        -    &    2\\
 20/04/91 &    48366.64 &     -   & 19.34(0.03) &  18.51(0.02) & 18.25(0.03) &        -    &    3\\
\noalign{\smallskip}
\hline
\noalign{\smallskip}
\multicolumn{8}{c}{SN 1991T continues}\\
\noalign{\smallskip}
16/04/91 &48362.85 &12.26(0.02) & 12.98(0.02)  &12.84(0.02) & 12.67(0.02) &12.87(0.02) &   10\\
17/04/91 &48363.80 &12.04(0.02) & 12.80(0.02)  &12.74(0.02) &       -    &       -   &   10\\
17/04/91 &48363.65 &      -   & 12.73(0.01)  &12.61(0.01) &       -    &       -   &   2\\
18/04/91 &48364.64 &      -   & 12.46(0.01)  &12.34(0.01) & 12.32(0.01) &       -   &   2\\
20/04/91 &48366.67 &      -   & 12.06(0.01)  &12.00(0.01) & 11.97(0.01) &       -   &   3\\
21/04/91 &48367.65 &      -   &        -    &11.89(0.01) &       -    &       -   &   3\\
22/04/91 &48368.70 &      -   & 11.78(0.01)  &11.80(0.01) & 11.70(0.01) &       -   &   3\\
20/05/91 &48396.56 &      -   & 13.39(0.01)  &12.32(0.01) & 12.32(0.01) &       -   &   3\\
20/05/91 &48396.56 &      -   & 13.48(0.01)  &12.29(0.01) &       -    &       -   &   3\\
20/05/91 &48396.56 &      -   &       -     &12.45(0.01) &       -    &       -   &   3\\
20/06/91 &48427.67 &      -   & 14.77(0.01)  &13.60(0.01) & 13.51(0.01) &       -   &   3  \\
20/06/91 &48427.67 &      -   &       -     &13.66(0.01) & 13.45(0.01) &       -   &   3   \\
20/06/91 &48427.67 &      -   &       -     &13.59(0.01) &       -    &       -   &   3  \\
01/08/91 &48469.51 &      -   & 15.42(0.02)  &      -    & 14.78(0.02) &       -   &   2\\
12/08/91 &48480.49 &      -   &       -     &15.29(0.02) &       -    &       -   &   2\\
02/12/91 &48593.46 &      -   & 16.98(0.06)  &17.01(0.05) & 17.62(0.06) &       -   &   7 \\
02/12/91 &48593.46 &      -   & 16.95(0.06)  &17.01(0.05) & 17.59(0.06) &       -   &   7 \\
03/12/91 &48594.46 &      -   & 17.06(0.06)  &17.10(0.06) & 17.65(0.06) &       -   &   7 \\
12/12/91 &48603.46 &      -   & 17.04(0.06)  &17.07(0.06) & 17.68(0.06) &       -   &   7 \\
\noalign{\smallskip}
\hline
\hline
\end{tabular}
\end{center}
\end{table*}

\begin{table*}
\begin{center}
\begin{tabular}{l l c c c c c l}
\multicolumn{8}{c}{{\bf Table A1.}}\\
\noalign{\smallskip}
\hline
\hline
\noalign{\smallskip}
Date & Julian Day & $U$ & $B$ & $V$ & $R$ & $I$ & telescope  \\
     & {\footnotesize +2,400,000} &     &     &     &     &     &             \\
\noalign{\smallskip}
\hline
\noalign{\smallskip}
\noalign{\smallskip}
\multicolumn{8}{c}{SN 1991T }\\
\noalign{\smallskip}
01/02/92 &48653.90 &      -   & 17.66(0.05)  &17.95(0.05) & 18.54(0.05) &       -   &   2\\
05/02/92 &48657.82 &      -   & 17.74(0.04)  &17.96(0.03) & 18.56(0.04) &       -   &   3\\
05/02/92 &48657.60 &      -   &       -     &17.88(0.07) &       -    &       -   &   7\\
06/02/92 &48658.66 &      -   & 17.75(0.07)  &17.94(0.08) & 18.52(0.06) &       -   &   7\\
29/02/92 &48681.78 &      -   & 18.02(0.06)  &18.34(0.05) & 18.97(0.05) &       -   &   2\\
04/03/92 &48686.46 &      -   & 18.25(0.10)  &18.27(0.06) & 19.00(0.08) &       -   &   7 \\
06/03/92 &48688.46 &      -   & 18.25(0.10)  &18.35(0.06) & 19.05(0.08) &       -   &   7 \\
11/03/92 &48692.67 &      -   & 18.42(0.07)  &18.42(0.07) & 19.07(0.06) &       -   &   6 \\
04/05/92 &48746.69 &      -   & 18.99(0.05)  &19.12(0.05) & 19.58(0.05) &19.00(0.08) &   3 \\
26/07/92 &48829.67 &      -   & 19.97(0.10)  &20.27(0.05) & 20.57(0.05) &       -   &   2 \\
23/12/92 &48979.67 &      -   & 21.04(0.10)  &20.95(0.08) &       -    &       -   &   3 \\
22/01/93 &49010.67 &      -   & 21.02(0.10)  &21.00(0.08) & 21.28(0.09) &       -   &   3 \\
17/02/93 &49035.67 &      -   &      -      &20.97(0.09) & 21.25(0.07) &       -   &   6 \\
27/02/93 &49045.79 &      -   & 21.08(0.10)  &20.99(0.09) &       -    &       -   &   2 \\
28/02/93 &49046.79 &      -   & 21.21(0.10)  &21.02(0.10) & 21.36(0.10) &       -   &   2\\
15/03/93 &49061.67 &      -   & 21.03(0.10)  &21.08(0.07) & 21.17(0.07) &       -   &   3 \\
26/04/93 &49103.60 &      -   & 21.06(0.10)  &     -     & 21.22(0.07) &       -   &   3 \\
27/04/93 &49104.60 &      -   & 21.06(0.10)  &21.21(0.08) &       -    &       -   &   3\\
17/03/94 &49428.78 &      -   & 21.43(0.11)  &21.45(0.10) &       -    &       -   &   6 \\
10/05/94 &49482.62 &      -   & 21.13(0.11)  &      -    &       -    &       -   &   3\\
15/05/94 &49487.65 &      -   & 21.30(0.11)  &      -    &       -    &       -   &   2 \\
17/03/96 &50159.77 &      -   & 21.49(0.12)  &21.48(0.10) &       -    &       -   &   3 \\
\noalign{\smallskip}
\hline
\noalign{\smallskip}
\multicolumn{8}{c}{SN 1992A }\\
\noalign{\smallskip}
14/01/92 &48635.06&      -      & 12.84(0.01) &   12.81(0.01) &   12.76(0.01) &        -  &    6\\
15/01/92 &48636.06&      -      & 12.86(0.01) &   12.79(0.01) &   12.75(0.01) &        -  &    6\\
16/01/92 &48637.03&      -      & 12.75(0.01) &   12.73(0.01) &        -     &        -  &    6 \\
17/01/92 &48638.03&      -      & 12.68(0.01) &   12.69(0.01) &        -     &        -  &    6 \\
18/01/92 &48639.03&      -      & 12.62(0.01) &   12.66(0.01) &   12.70(0.01) &        -  &    6 \\
04/02/92 &48656.57&      -      & 14.17(0.01) &   13.43(0.01) &   13.32(0.01) &        -  &    3 \\ 
11/03/92 &48692.03&      -      & 16.03(0.03) &   15.19(0.02) &   15.10(0.02) &        -  &    3 \\  
26/07/92 &48829.92&      -      & 17.94(0.03) &   18.24(0.03) &   18.53(0.03) &        -  &    2 \\   
01/09/92 &48867.80 &      -     & 18.43(0.03) &   18.61(0.03) &   19.15(0.04) &  18.90(0.05)&   3\\ 
30/09/92 &48895.40&      -      &      -     &        -     &   20.00(0.07) &        -  &    5      \\
23/10/92 &48918.74&      -      & 18.94(0.05) &   19.60(0.05) &   20.18(0.05) &  19.69(0.08)&    2 \\   
29/11/92 &48956.70&      -      & 19.69(0.08) &   19.98(0.05) &   20.66(0.05) &  20.09(0.08)&    3 \\
02/12/92 &48958.63&      -      &      -     &   20.08(0.07) &        -     &        -  &    6       \\
22/12/92 &48979.64&      -      & 19.99(0.08) &   20.29(0.07) &   20.76(0.06) &  20.16(0.07)&    3   \\   
24/01/93 &49011.55&      -      & 20.43(0.08) &   20.68(0.09) &   21.28(0.08) &  20.54(0.10)&    3   \\        
24/01/93 &49011.55&      -      &      -     &   20.76(0.09) &   21.15(0.07) &        -  &    3   \\        
27/02/93 &49045.54&      -      & 20.74(0.10) &   21.23(0.12) &        -     &  21.10(0.15)&    2 \\   
28/02/93 &49046.50&      -      &      -     &   21.22(0.11) &        -     &        -  &    2 \\   
27/03/93 &49073.54&      -      & 21.41(0.07) &   21.49(0.07) &   21.82(0.07) &        -  &    6   \\
07/07/93 &49176.20&      -      & $>$22.5 &        -     &        -     &        -  &         1\\
02/08/94 &49566.50&      -      &      -     &   25.9 (0.30) &        -     &        -  &     11  \\
04/09/96 &50331.12&      -      &      -     &$>$27.0     &        -     &        -  &       11 \\

\noalign{\smallskip}
\hline
\noalign{\smallskip}
\multicolumn{8}{c}{SN 1992K }\\
\noalign{\smallskip}
  07/04/92  &    48719.80 &     -   & 19.23(0.03) & 18.18(0.02) &  17.90(0.02) & 17.51(0.02) &      2  \\
\noalign{\smallskip}
\hline
\noalign{\smallskip}
\multicolumn{8}{c}{SN 1993H }\\
\noalign{\smallskip}
  26/04/1993  &49103.67  &    -    &20.12(0.04)   &19.02(0.03)  & 18.51(0.03)   &      -    &    3\\
  05/05/1993  &49112.90  &    -    &      -      &       -    & 18.85(0.05)   &      -    &    1\\
  16/05/1993  &49123.67  &    -    &20.53(0.13)   &19.58(0.07)  & 19.32(0.06)   &      -    &    1\\
  17/05/1993  &49124.63  &    -    &20.49(0.04)   &19.70(0.05)  & 19.46(0.04)   &      -    &    3 \\
  08/07/1993  &49176.67  &    -    &21.73(0.20)   &21.04(0.10)  & 21.18(0.15)   &      -    &    1\\
\noalign{\smallskip}
\hline
\hline
\end{tabular}
%\caption{ESO-ASIAGO SN ARCHIVE DATA }
\end{center}
\end{table*}

\begin{table*}
\begin{center}
\begin{tabular}{l l c c c c c l}
\multicolumn{8}{c}{{\bf Table A1.}}\\
\noalign{\smallskip}
\hline
\hline
\noalign{\smallskip}
Date & Julian Day & $U$ & $B$ & $V$ & $R$ & $I$ & telescope  \\
     & {\footnotesize +2,400,000} &     &     &     &     &     &             \\
\noalign{\smallskip}
\hline
\noalign{\smallskip}
\multicolumn{8}{c}{SN 1993L }\\
\noalign{\smallskip}
 05/05/93&  49112.90  &     -  & 14.84(0.05)& 13.97(0.03)& 13.86(0.03)&       -  &   1  \\
 16/05/93&  49123.83  &     -  & 15.94(0.09)& 14.60(0.04)& 14.11(0.05)&       -  &   1  \\
 16/05/93&  49123.84  &     -  & 15.94(0.09)& 14.60(0.04)& 14.04(0.05)&       -  &   1  \\
 17/05/93&  49124.88  &     -  &       -   &       -   & 14.17(0.02)&       -  &   3 \\
 27/05/93&  49134.40  &     -  & 16.75(0.04)& 15.23(0.01)& 14.87(0.02)& 14.51(0.01)&   6\\
 30/05/93&  49137.40  &     -  &       -   & 15.30(0.01)& 14.79(0.02)&       -  &   6    \\
 14/06/93&  49152.70  &     -  & 16.89(0.07)& 15.81(0.03)& 15.44(0.03)&       -  &   5 \\
 21/06/93&  49159.70  &     -  & 16.93(0.07)& 15.88(0.03)& 15.60(0.04)&       -  &   5 \\
 01/07/93&  49169.43  &     -  & 17.02(0.04)& 16.21(0.02)& 16.05(0.02)& 16.01(0.02)&   6    \\
 07/07/93&  49176.15  &     -  & 17.29(0.10)& 16.42(0.06)& 16.29(0.05)&       -  &   1  \\
 10/07/93&  49178.40  &     -  & 17.31(0.04)& 16.52(0.02)& 16.35(0.02)& 16.32(0.03)&   3  \\    
 11/07/93&  49179.40  &     -  & 17.30(0.04)& 16.56(0.02)& 16.41(0.03)& 16.34(0.03)&   3  \\
 13/07/93&  49181.40  &     -  & 17.38(0.04)& 16.59(0.02)& 16.44(0.03)& 16.45(0.03)&   3  \\    
 23/08/93&  49223.84  &     -   & 17.81(0.05)& 17.55(0.03)& 17.68(0.04)&       -   &  3  \\
 10/09/93&  49241.70  &19.40(0.09)& 18.22(0.09)& 18.00(0.05)& 18.19(0.05)& 18.19(0.05)&  5  \\   
 18/11/93&  49309.52  &     -   & 19.05(0.12)& 19.06(0.06)& 19.36(0.06)& 19.16(0.07)&  5  \\   
 13/12/94&  49332.57  &     -   &       -   & 19.70(0.07)& 20.06(0.07)&       -   &  5 \\
 15/12/94&  49336.54  &     -   &       -   & 19.93(0.06)& 19.92(0.05)&       -   &  3  \\
 14/01/94&  49367.45  &     -   & 19.64(0.09)& 20.26(0.10)&       -   &       -   &  3\\
 06/05/94&  49478.67  &     -   & 21.49(0.20)&       -   &       -   &       -   &  5 \\
 09/05/94&  49481.83  &     -   & 21.38(0.15)& 21.51(0.10)&       -   &       -   &  3   \\      
 14/05/94&  49486.85  &     -   & 21.45(0.20)& 21.52(0.20)&       -   &       -   &  2 \\
 04/06/94&  49507.79  &     -   & 21.73(0.22)& 21.73(0.12)&       -   &       -   &  3    \\     
 06/06/94&  49509.88  &     -   & 21.72(0.20)& 21.94(0.30)& 21.70(0.20)&       -   &  2  \\    
 07/09/94&  49602.63  &     -   & 22.45(0.40)& 22.47(0.40)&       -   &       -   &  5\\
 07/09/94&  49602.63  &     -   &       -   &       -   &$>$21.91&$>$21.3       &  5\\
 08/09/94&  49603.58  &     -   & 22.88(0.40)& 22.59(0.40)&       -   &       -   &  5\\
 08/09/94&  49603.58  &     -   &       -   &       -   &$>$22.02&$>$21.41      &  5\\
 28/09/94&  49624.67  &     -   & 23.00(0.30)&       -   &       -   &       -   &  3\\ 
 29/05/95&  49866.67  &     -   &$>$22.60&       -   &       -   &$>$21.5       &  6 \\
\noalign{\smallskip}
\hline
\noalign{\smallskip}
\multicolumn{8}{c}{SN 1994D}\\
\noalign{\smallskip}
07/06/94  &49510.54 &     -   & 15.67(0.02)& 15.25(0.02)& 15.14(0.02)& 15.40(0.03)&   3\\
08/06/94  &49512.46 &     -   & 15.64(0.02)& 15.24(0.02)& 15.24(0.02)& 15.44(0.03)&   3 \\
09/06/94  &49513.46 &     -   &      -   & 15.35(0.02)& 15.34(0.02)&       -   &   3\\
13/06/94  &49516.51 &16.25(0.02)& 15.75(0.01)& 15.43(0.01)& 15.42(0.01)& 15.58(0.01)&   2 \\
14/06/94  &49517.60 &     -   &      -  &      -    & 15.47(0.01)&       -   &   2  \\
25/06/94  &49529.47 &     -   & 15.99(0.03)& 15.69(0.03)& 15.79(0.03)& 16.12(0.03)&   5\\
13/07/94  &49546.51 &     -   & 16.26(0.03)& 16.14(0.04)& 16.20(0.03)& 16.59(0.04)&   5 \\
08/01/95  &49725.78 &     -   & 18.62(0.04)& 19.03(0.04)&      -   &       -   &   3\\
28/01/95  &49745.85 &     -   & 19.11(0.09)& 19.47(0.06)&      -   &       -   &   5  \\
04/02/95  &49752.85 &     -   & 19.07(0.03)&      -   &      -   &       -   &   2    \\
05/02/95  &49753.79 &     -   & 19.03(0.03)& 19.60(0.02)&      -   &       -   &   2    \\
20/02/95  &49768.87 &     -   & 19.46(0.10)& 19.97(0.10)&      -   &       -   &   1\\
27/02/95  &49776.30 &     -   & 19.75(0.08)&      -   &      -   &       -   &   6       \\
28/02/95  &49776.46 &     -   & 19.79(0.15)&      -   &      -   &       -   &   7\\
24/04/95  &49831.67 &     -   & 20.56(0.14)&      -   &      -   &       -   &   5 \\
24/04/95  &49831.67 &     -   &      -   &$>$20.50&     -   &       -   &        5 \\
28/05/95  &49865.67 &     -   & 20.91(0.20)&      -   &      -   &       -   &     6 \\
28/05/95  &49865.67 &     -   &      -   &$>$20.54&     -   &       -   &        6 \\
\noalign{\smallskip}
\hline
\noalign{\smallskip}
\multicolumn{8}{c}{SN 1994M }\\
\noalign{\smallskip}
  14/05/94 &49486.56  &    -    & 17.28(0.01) &16.84(0.01)& 16.83(0.01) & 17.25(0.01) &    2\\
\noalign{\smallskip}
\hline
\hline
\end{tabular}
%\caption{ESO-ASIAGO SN ARCHIVE DATA }
\end{center}
\end{table*}

\begin{table*}
\begin{center}
\begin{tabular}{l l c c c c c l}
\multicolumn{8}{c}{{\bf Table A1.}}\\
\noalign{\smallskip}
\hline
\hline
\noalign{\smallskip}
Date & Julian Day & $U$ & $B$ & $V$ & $R$ & $I$ & telescope  \\
     & {\footnotesize +2,400,000} &     &     &     &     &     &             \\
\noalign{\smallskip}
\hline
\noalign{\smallskip}
\multicolumn{8}{c}{SN 1994ae }\\
\noalign{\smallskip}
  23/12/94 &49709.85  &     -   & 15.02(0.05) &14.20(0.03)& 13.94(0.03)& 13.96(0.03) &   1\\
  13/01/95 &49730.33  &     -   & 16.14(0.01) &15.17(0.01)& 14.89(0.01)& 14.67(0.01) &   4\\
  28/01/95 &49745.77  &     -   & 16.34(0.04) &15.60(0.03)& 15.40(0.03)& 15.33(0.03) &   5\\
  05/02/95 &49753.70  &     -   & 16.53(0.02) &15.83(0.02)& 15.74(0.02)& 15.71(0.02) &   2\\
  20/02/95 &49768.79  &     -   & 16.69(0.05) &16.20(0.05)& 16.12(0.05)& 16.18(0.04) &   1\\
  27/02/95 &49776.27  &     -   &      -    &16.38(0.02)&       -   &       -    &   6\\         
  27/02/95 &49775.46  &     -   & 16.90(0.07) &16.28(0.05)& 16.26(0.05)& 16.56(0.04) &   7\\
  28/02/95 &49776.46  &     -   & 16.90(0.07) &16.30(0.05)& 16.37(0.06)& 16.63(0.05) &   7\\ 
  31/03/95 &49807.54  &     -   &     -     &       -  &       -   & 17.35(0.04) &   3\\ 
  20/04/95 &49827.68  &     -   &     -     &17.39(0.07)& 17.63(0.07& 18.00(0.06) &   1\\
  24/04/95 &49831.67  &     -   & 17.80(0.06) &17.77(0.05)& 17.89(0.05)&       -    &   5\\
  29/05/95 &49866.67  &     -   & 18.46(0.07) &18.39(0.07)& 18.82(0.05)&       -    &   6 \\
  19/01/96 &50101.77  &     -   & 21.45(0.20) &21.65(0.15)&       -   &       -    &   2  \\      
  21/02/96 &50135.67  &     -   & 21.89(0.30) &22.09(0.17)&       -  &        -    &   5 \\ 
  13/05/96 &50216.67  &     -   &$>$22.70&$>$22.80&      -  &        -          &   5 \\
\noalign{\smallskip}
\hline
\noalign{\smallskip}
\multicolumn{8}{c}{SN 1995D }\\
\noalign{\smallskip}
  20/02/95& 49768.78 &       -   &13.39(0.01)& 13.44(0.01)& 13.47(0.02)& 13.68(0.03) &   1\\
  27/02/95& 49776.30 &       -   &     -   & 13.64(0.01)& 13.75(0.01)&       -   &   6   \\ 
  30/03/95& 49806.59 &       -   &     -   & 15.24(0.01)&      -   &       -   &   3    \\
  19/04/95& 49827.68 &       -   &16.74(0.12)& 15.82(0.10)& 15.64(0.10)& 15.59(0.10) &   1  \\
  25/04/95& 49831.67 &       -   &16.80(0.12)& 15.97(0.11)& 15.93(0.10)& 15.71(0.10) &   5      \\
  29/05/95& 49867.67 &       -   &17.40(0.06)& 16.87(0.06)& 16.88(0.04)& 17.10(0.05) &   6   \\
  22/11/95& 50044.62 &       -   &19.86(0.15)& 20.13(0.10)&      -   &       -   &   7  \\
  23/11/95& 50045.58 &       -   &19.65(0.14)&      -   &      -   &       -   &   7 \\
  25/12/95& 50077.85 &       -   &20.08(0.10)& 20.48(0.10)&      -   &       -   &   2   \\
  19/01/96& 50101.67 &       -   &20.79(0.22)& 20.84(0.12)& 21.69(0.15)& 21.10(0.20) &   5  \\ 
  01/02/96& 50114.67 &       -   &21.11(0.09)& 21.50(0.08)&      -   &       -   &   1  \\ 
  01/02/96& 50114.67 &       -   &     -   & 21.32(0.08)&      -   &       -   &   1   \\ 
  15/02/96& 50128.56 &       -   &21.13(0.20)& 21.24(0.20)&      -   &       -   &   7\\
  16/02/96& 50129.60 &       -   &     -   &      -   & 22.26(0.30)&      -    &   7 \\
  16/02/96& 50129.57 &       -   &     -   & 21.24(0.20)&      -   &       -   &   7\\
  18/02/96& 50131.68 &       -   &20.86(0.25)& 21.27(0.10)& 22.12(0.10)&       -   &   2 \\
  21/02/96& 50134.67 &       -   &21.31(0.30)& 21.36(0.15)& 22.15(0.20)& 21.04(0.10) &   5 \\  
  19/04/96& 50192.67 &       -   &22.16(0.50)& 22.30(0.22)& 23.39(0.50)&       -   &   5   \\
  10/02/97& 50489.67 &       -   &     -   &$>$23.0&      -   &       -   &      3\\
\noalign{\smallskip}
\hline
\noalign{\smallskip}
\multicolumn{8}{c}{SN 1995ac }\\
\noalign{\smallskip}
02/10/95   &49992.67 &      -   & 17.24(0.02) & 17.26(0.02) & 17.19(0.02)&  17.32(0.02)&    3 \\
13/10/95   &50004.64 &      -   & 17.93(0.15) & 17.53(0.06) & 17.50(0.06)&  17.79(0.15)&    3\\
14/11/95   &50035.67 &      -   & 20.25(0.08) & 19.19(0.07) & 18.59(0.06)&  18.47(0.07)&    6  \\
09/06/96   &50243.83 &      -   &$>$23.49& 22.79(0.20) &      -  &        -   &       3 \\
\noalign{\smallskip}
\hline
\noalign{\smallskip}
\multicolumn{8}{c}{SN 1995bd }\\
\noalign{\smallskip}
18/01/96 & 50100.67   &    -    &18.19(0.01)   &17.12(0.01)   &     -   &    -   &    2 \\
\noalign{\smallskip}
\hline
\noalign{\smallskip}
\multicolumn{8}{c}{SN 1996bo }\\
\noalign{\smallskip}
24/10/96 &50380.65  &     -    &16.33(0.03) & 16.05(0.03) & 15.79(0.03) &     -     &    5 \\
19/11/96 &50406.64  &18.46(0.08 &17.81(0.08) & 16.60(0.04) & 16.21(0.03) & 16.09(0.04) &    5 \\
\noalign{\smallskip} 
\hline
\hline
\end{tabular}
%\caption{ESO-ASIAGO SN ARCHIVE DATA }
%\label{esoasiagoarchive}
\end{center}
\end{table*}

\begin{table}
\begin{center}
\begin{tabular}{l l c c c c c l}
\multicolumn{8}{c}{{\bf Table A1.}}\\
\noalign{\smallskip}
\hline
\hline
\noalign{\smallskip}
Date & Julian Day & $U$ & $B$ & $V$ & $R$ & $I$ & telescope  \\
     & {\footnotesize +2,400,000} &     &     &     &     &     &             \\
\noalign{\smallskip}
\hline
\multicolumn{8}{c}{SN 1997bp }\\
\noalign{\smallskip}
08/04/97 &50547.49&       -   & 14.19(0.02)& 14.02(0.02) &13.91(0.02)&       -    &   7\\
17/04/97 &50555.73& 14.92(0.04)& 14.47(0.04)& 14.01(0.03) &13.99(0.03)& 14.47(0.03) &   5\\
18/04/97 &50556.68& 15.06(0.04)& 14.54(0.04)& 14.01(0.03) &14.01(0.03)& 14.50(0.03) &   5\\
19/04/97 &50557.65& 15.10(0.05)& 14.64(0.05)& 14.12(0.03) &14.12(0.03)& 14.62(0.03) &   5\\
26/04/97 &50565.74& 16.06(0.06)& 15.47(0.05)& 14.58(0.03) &14.43(0.03)& 14.81(0.03) &   5\\
13/05/97 &50581.67&       -   & 16.93(0.01)& 15.43(0.01) &14.93(0.01)& 14.59(0.01) &   2  \\
02/01/98 &50815.88&       -   & 20.11(0.14)& 20.07(0.08) &20.79(0.09)& 20.44(0.09) &   5\\
02/02/98 &50846.81&       -   & 20.66(0.15)& 20.65(0.10) &     -    &      -     &   8\\
26/03/98 &50898.75&       -   &      -    &      -     &21.85(0.20)&      -     &   5\\
29/05/98 &50962.64&       -   & 22.17(0.20)& 22.16(0.15) &     -    &      -     &   8\\
\noalign{\smallskip}
\hline
\noalign{\smallskip}
\multicolumn{8}{c}{SN 1997br }\\
\noalign{\smallskip}
  17/04/97  &50555.75& 13.58(0.03) &14.18(0.03)& 13.93(0.03)& 13.74(0.03)& 13.65(0.03)&  5 \\
  19/04/97  &50557.66& 13.64(0.03) &14.13(0.03)& 13.87(0.03)& 13.70(0.03)& 13.62(0.03)&  5 \\
  13/05/97  &50581.77&      -    &16.03(0.01)& 14.64(0.01)& 14.25(0.01)&       -  &  2\\
  09/08/97  &50670.38&      -    &     -   &      -   & 17.24(0.05)&       -  &  1\\
  22/03/98  &50894.79&      -    &21.26(0.18)& 21.06(0.15)& 21.63(0.17)&       -  &  1\\
  29/05/98  &50962.67&      -    &21.94(0.20)& 21.82(0.15)& 22.04(0.19)&       -  &  8\\
\noalign{\smallskip}
\hline
\noalign{\smallskip}
\multicolumn{8}{c}{SN 1999aa }\\
\noalign{\smallskip}
16/03/99&   51254.67 &   16.86(0.07)  &16.51(0.06) & 15.76(0.02)  & 15.66(0.03) & 15.86(0.04)& 5  \\
17/03/99&   51255.57 &   16.91(0.07)  &16.61(0.06) & 15.80(0.03)  & 15.67(0.03) & 15.84(0.04)& 5  \\
12/04/99&   51281.46 &   18.27(0.05)  &17.98(0.05) & 17.10(0.04)  & 16.80(0.04) & 16.73(0.05)& 8 \\
\noalign{\smallskip}
\hline
\noalign{\smallskip}
\multicolumn{8}{c}{SN 1999dk }\\
\noalign{\smallskip}
25/08/99  & 51415.67   & 14.94(0.07)  & 15.06(0.05)  & 14.93(0.04)  & 14.92(0.04)  & 15.30(0.04)   & 1 \\
02/09/99  & 51423.67   & 15.62(0.07)  & 15.45(0.06)  & 15.14(0.04)  & 15.23(0.05)  & 15.78(0.05)   & 1 \\
05/09/99  & 51426.67   & 16.12(0.08)  & 15.82(0.07)  & 15.38(0.04)  & 15.49(0.05)  & 15.95(0.05)   & 1 \\
14/09/99  & 51435.78   & 17.24(0.15)  & 16.75(0.05)  & 15.85(0.03)  & 15.68(0.03)  & 15.92(0.03)   & 8 \\
17/09/99  & 51438.67   & 17.54(0.20)  & 17.11(0.05)  & 16.01(0.03)  & 15.70(0.03)  & 15.80(0.03)   & 8 \\
08/10/99  & 51459.56   & 19.10(0.15)  & 18.12(0.03)  & 16.63(0.03)  & 16.58(0.02)  & 16.63(0.02)   & 9\\
\noalign{\smallskip}
\hline
\noalign{\smallskip}
\multicolumn{8}{c}{SN 2000cx }\\
\noalign{\smallskip}
27/07/00  &51753.90 &   -         & 13.44(0.02) & 13.24(0.03) & 13.35(0.03) & 13.77(0.02)  &  6\\
18/11/00 &51866.55 & 18.31(0.02)  & 17.86(0.01) & 17.76(0.01) & 17.80(0.02) & 18.21(0.02)  &  1 \\
\noalign{\smallskip}
\hline
\hline
\end{tabular}
%\caption{ESO-ASIAGO SN ARCHIVE DATA }
%\label{esoasiagoarchive}
\end{center}
\end{table}

%%%%%%%%%%%%%%%%%%%%%%%%%%%%%%%%%%%%%%%%%%%%%%%%%%%%%%%%%%%%%%%%%%%%%%%%%%%%%%%%
\onecolumn

\section{SN\protect\lowercase{e} data}\label{Asnedata}
Table~\ref{tab11}: 
(1) the SN name; (2) the apparent B magnitude at maximum; (3) the
stretch factor ($s^{-1}$) and (4) the $\Delta m_{15}(B)_{obs}$
measured as explained in \S \ref{analysis}, not corrected for colour
excess; (5) the Galactic colour excess $E(B-V)_{gal}$ from
\cite{schlegel}; (6) the colour excess at the blue maximum
$E(B-V)_{max}$ corrected for galactic extinction; (7) the colour
excess measured in the tail: $E(B-V)_{tail}$, corrected for galactic extinction;
(8) the weighted mean of $E(B-V)_{max}$ and $E(B-V)_{tail}$:
$E(B-V)_{host}$. $E(B-V)_{host}=E(B-V)_{tail}$ when $E(B-V)_{max}$
it is not available; (9) distance modulus $\mu$ from the Nearby Galaxy
Catalogue \citep{tully88}; (10) the recession velocity $v_{3k}$ with
respect to the microwave background from NED.
%(* marks objects whose $v_{3k}$  velocity is not available in RC3;
%** from \cite{hamuy96a}).  
The data sources are reported in Table~\ref{tab22}.
\\
\floatplacement{table}{h}
\begin{table*}
\begin{center}
\caption{}
\label{tab11}
\setlength\tabcolsep{2pt}
\begin{tabular}[b]{l c c c c c c c c  r }
\hline
SN& m(B) & $s^{-1}$ &  $\Delta m_{15}(B)_{obs}$  & $E(B-V)_{gal}$ 
 & $E(B-V)_{max}$ & $E(B-V)_{tail}$ & $E(B-V)_{host}$& $\mu_{tully}$ & $v_{3k}$ \\
(1)& (2) & (3) &  (4)  & (5) 
 & (6) & (7) & (8)& (9) & (10)  \\
\noalign{\smallskip}
\hline
\hline
1937C &  8.94(0.30)  &      -     &  0.85(0.10) & 0.014 &  0.08(0.20) &  0.06(0.05) &  0.06(0.05)  & 28.22 &   543   \\ 
1972E &  8.40(0.04)  &  0.96(0.06) &  1.05(0.05) & 0.056 & -0.04(0.10) &  0.03(0.05) &  0.01(0.05)  & 27.53 &   671   \\ 
1972J & 14.80(0.20)  &  1.22(0.09) &  1.45(0.20) & 0.046 & -0.05(0.11) &  0.00(0.20) &  0.00(0.05)  &      - &  2865   \\ 
1974G & 12.34(0.04)  &  1.14(0.08) &  1.40(0.04) & 0.019 &  0.25(0.20) &  0.25(0.20) &  0.25(0.05)  & 29.93 &   992   \\ 
1980N & 12.49(0.04)  &  1.06(0.07) &  1.30(0.05) & 0.021 &  0.05(0.06) &  0.12(0.05) &  0.09(0.05)  & 31.14 &  1678   \\ 
1981B & 12.04(0.04)  &  1.10(0.07) &  1.13(0.05) & 0.018 &  0.12(0.06) &  0.10(0.14) &  0.11(0.05)  &      - &  2141   \\ 
1982B & 13.65(0.20)  &      -     &  0.94(0.10) & 0.064 &       -     &  0.14(0.14) &  0.14(0.14)  & 32.68 &  2203   \\ 
1983G & 13.00(0.10)  &  1.18(0.08) &  1.30(0.20) & 0.034 &  0.19(0.11) &  0.42(0.05) &  0.35(0.16)  & 30.89 &  1566   \\ 
1986G & 12.48(0.04)  &  1.37(0.11) &  1.69(0.05) & 0.115 &  0.95(0.08) &  0.67(0.05) &  0.78(0.20)  & 28.45 &  806\\%*   \\ 
1989B & 12.38(0.12)  &  1.04(0.07) &  1.28(0.05) & 0.032 &  0.41(0.06) &  0.46(0.14) &  0.42(0.05)  & 29.10 &  1067   \\ 
1990N & 12.75(0.04)  &  1.04(0.07) &  1.05(0.05) & 0.026 &  0.09(0.06) &  0.18(0.05) &  0.14(0.06)  & 31.13 &  1322   \\ 
1990O & 16.59(0.04)  &  0.91(0.08) &  0.94(0.05) & 0.093 &  0.05(0.10) &  0.10(0.05) &  0.08(0.05)  &      - & 9175\\%*\\ 
1990T & 17.27(0.20)  &      -     &  1.13(0.20) & 0.053 &  0.08(0.30) &  0.16(0.05) &  0.15(0.06)  &      - &  11893\\%*\\ 
1990Y & 17.69(0.20)  &      -     &  1.13(0.20) & 0.008 &  0.26(0.30) &  0.34(0.10) &  0.32(0.05)  &      - &  11652\\%* \\ 
1991S & 17.78(0.05)  &      -     &  1.00(0.10) & 0.026 &  0.06(0.30) &  0.13(0.10) &  0.11(0.05)  &      - &  16807\\%*\\ 
1991T & 11.69(0.04)  &  0.93(0.05) &  0.94(0.05) & 0.022 &  0.21(0.06) &  0.19(0.04) &  0.20(0.05)  & 30.65 &  2070   \\ 
1991U & 16.67(0.10)  &      -     &  1.11(0.10) & 0.062 &  0.07(0.30) &  0.15(0.20) &  0.12(0.05)  &      - &  9724\\%*\\ 
1991ag & 14.67(0.04)  &  0.89(0.02) &  0.87(0.05) & 0.062 &  0.06(0.30) &  0.08(0.05) &  0.08(0.05)  &      - & 4521\\%* \\ 
1991bg & 14.75(0.04)  &  1.47(0.22) &  1.94(0.05) & 0.041 & -0.00(0.07) &  0.00(0.09) &  0.00(0.05)  & 31.13 &  1282   \\ 
1992A & 12.56(0.04)  &  1.25(0.09) &  1.47(0.05) & 0.018 &  0.05(0.07) &  0.02(0.05) &  0.04(0.05)  & 31.14 &  1737   \\ 
1992J & 17.88(0.20)  &      -     &  1.69(0.20) & 0.057 &       -     &  0.10(0.08) &  0.10(0.08)  &      - & 13828\\%*   \\ 
1992K & 16.31(0.20)  &      -     &  1.94(0.20) & 0.101 &       -     &  0.00(0.05) &  0.00(0.05)  &      - &  3324   \\ 
1992P & 16.14(0.20)  &  1.00(0.03) &  1.05(0.30) & 0.021 &  0.05(0.06) &  0.12(0.05) &  0.09(0.05)  &      - &  7939   \\ 
1992ae & 18.62(0.04)  &      -     &  1.30(0.10) & 0.036 &  0.12(0.11) &  0.05(0.10) &  0.09(0.05)  &      - & 22442\\%*   \\ 
1992ag & 16.64(0.20)  &  1.07(0.07) &  1.20(0.20) & 0.097 &  0.15(0.06) &  0.45(0.20) &  0.22(0.21)  &      - & 8095\\%* \\ 
1992al & 14.59(0.04)  &  1.00(0.03) &  1.09(0.05) & 0.034 & -0.01(0.06) &  0.07(0.04) &  0.04(0.05)  &      - &  4214   \\ 
1992aq & 19.42(0.10)  &      -     &  1.69(0.05) & 0.012 &  0.05(0.31) &  0.00(0.20) &  0.02(0.05)  &      - & 30014\\%*   \\ 
1992au & 18.17(0.20)  &      -     &  1.69(0.30) & 0.017 &       -     &  0.00(0.30) &  0.00(0.30)  &      - & 18225\\%*   \\ 
1992bc & 15.13(0.04)  &  0.93(0.05) &  0.90(0.05) & 0.022 &  0.01(0.06) & -0.03(0.05) &  0.00(0.05)  &      - & 5876\\%*   \\ 
1992bg & 17.39(0.04)  &  1.00(0.03) &  1.15(0.05) & 0.185 & -0.04(0.20) &  0.04(0.05) &  0.02(0.06)  &      - & 10936\\%* \\ 
1992bh & 17.68(0.04)  &  0.98(0.03) &  1.13(0.05) & 0.022 &  0.10(0.10) &  0.13(0.10) &  0.12(0.05)  &      - & 13519\\%*   \\ 
1992bk & 18.07(0.04)  &      -     &  1.67(0.05) & 0.015 &  0.04(0.21) & -0.01(0.10) &  0.01(0.05)  &      - &  17551\\%*   \\ 
1992bl & 17.34(0.04)  &  1.23(0.05) &  1.56(0.05) & 0.011 & -0.02(0.11) &  0.04(0.05) &  0.02(0.05)  &      - & 12661\\%*   \\ 
1992bo & 15.86(0.04)  &  1.33(0.11) &  1.69(0.05) & 0.027 & -0.02(0.08) & -0.01(0.12) &  0.00(0.05)  &      - &  5164\\%*   \\ 
1992bp & 18.53(0.04)  &  1.12(0.04) &  1.52(0.20) & 0.069 & -0.02(0.11) &  0.00(0.20) &  0.00(0.05)  &      - & 23557\\%*   \\ 
1992bs & 18.33(0.04)  &  1.01(0.03) &  1.15(0.05) & 0.011 &  0.06(0.10) &  0.05(0.10) &  0.05(0.05)  &      - & 18787\\%*  \\ 
1993B & 18.71(0.06)  &  0.99(0.03) &  1.30(0.05) & 0.079 & -0.02(0.20) &  0.24(0.10) &  0.15(0.18)  &      - & 21011\\%*   \\ 
1993H & 16.99(0.04)  &  1.18(0.04) &  1.69(0.10) & 0.060 &  0.16(0.08) &  0.06(0.05) &  0.10(0.07)  &      - &  7523   \\ 
1993L & 13.40(0.20)  &      -     &  1.47(0.30) & 0.014 &  0.31(0.21) &  0.32(0.11) &  0.32(0.05)  &      - &  1387   \\ 
1993O & 17.79(0.04)  &  1.05(0.07) &  1.26(0.05) & 0.053 & -0.07(0.06) &  0.04(0.08) &  0.00(0.08)  &      - & 15867\\%*   \\ 
1993ac & 18.45(0.20)  &      -     &  1.25(0.20) & 0.163 &       -     &  0.04(0.10) &  0.04(0.10)  &      - & 14674\\%*   \\ 
1993ae & 15.44(0.20)  &      -     &  1.47(0.20) & 0.038 & -0.01(0.30) &  0.00(0.05) &  0.00(0.05)  &      - &  5405   \\ 
1993ag & 18.30(0.04)  &  1.06(0.03) &  1.30(0.20) & 0.112 &  0.15(0.06) &  0.10(0.10) &  0.13(0.05)  &      - & 15031\\%**  \\ 
1993ah & 16.32(0.04)  &  1.19(0.14) &  1.45(0.10) & 0.020 &       -     &  0.15(0.10) &  0.15(0.10)  &      - &  8604   \\ 
1994D & 11.86(0.04)  &  1.27(0.10) &  1.31(0.05) & 0.022 & -0.02(0.06) &  0.05(0.08) &  0.01(0.05)  & 31.13 &   793   \\ 
1994M & 16.35(0.06)  &  1.18(0.04) &  1.45(0.20) & 0.023 &  0.12(0.11) &  0.18(0.05) &  0.16(0.05)  &      - & 7289\\%*   \\ 
1994Q & 16.44(0.10)  &      -     &  0.90(0.10) & 0.017 &       -     &  0.11(0.06) &  0.11(0.06)  &      - &  8954\\%*   \\ 
\noalign{\smallskip}
\hline
\end{tabular}
\end{center}
\end{table*}

%%%%%%%%\end{sidewaystable*} 
%%%%%%%%%\begin{sidewaystable*} 
\begin{table*}
\begin{center}
%\caption{}
%%%%%%%%%%%%%\renewcommand{\arraystretch}{1.4}
\setlength\tabcolsep{2pt}
\begin{tabular}[t]{l c c c c c c c c r}
\multicolumn{10}{c}{{\bf Table B1.}}\\
\noalign{\smallskip}
\hline
SN& m(B) & $s^{-1}$ &  $\Delta m_{15}(B)_{obs}$  & $E(B-V)_{gal}$ 
 & $E(B-V)_{max}$ & $E(B-V)_{tail}$ & $E(B-V)_{host}$& $\mu_{tully}$ & $v_{3k}$ \\
(1)& (2) & (3) &  (4)  & (5) 
 & (6) & (7) & (8)& (9) & (10)  \\
\noalign{\smallskip}
\hline
\hline 
1994T & 17.23(0.10)  &     -    &  1.45(0.05) & 0.029 &  0.06(0.30) &  0.07(0.20) &  0.07(0.05)  &      - & 10708\\%*   \\ 
1994ae & 13.15(0.06)  &  1.04(0.07) &  0.87(0.10) & 0.030 &  0.17(0.06) &  0.16(0.09) &  0.16(0.05)  & 31.85 &  1609   \\ 
1995D & 13.44(0.04)  &  0.99(0.06) &  1.03(0.05) & 0.058 &  0.07(0.06) &  0.13(0.07) &  0.10(0.05)  & 32.43 &  2289   \\ 
1995E & 16.82(0.04)  &  0.96(0.03) &  1.19(0.05) & 0.027 &  0.77(0.06) &  0.72(0.14) &  0.76(0.05)  &      - &  3510   \\ 
1995ac & 17.21(0.04)  &  0.96(0.06) &  0.95(0.05) & 0.042 &  0.06(0.06) & -0.01(0.05) &  0.02(0.05)  &      - & 14635\\%*   \\ 
1995ak & 16.09(0.10)  &  1.11(0.07) &  1.45(0.10) & 0.038 & -0.04(0.11) &  0.28(0.06) &  0.16(0.23)  &      - &  6589\\%*   \\ 
1995al & 13.36(0.04)  &  0.93(0.03) &  0.89(0.05) & 0.014 &  0.21(0.06) &  0.19(0.05) &  0.20(0.05)  & 32.00 &  1797   \\ 
1995bd & 17.27(0.04)  &  0.93(0.05) &  0.88(0.05) & 0.498 &  0.37(0.06) &  0.20(0.05) &  0.28(0.12)  &      - &  4326\\%*   \\ 
1996C & 16.59(0.08)  &  0.87(0.02) &  0.94(0.10) & 0.013 &  0.11(0.20) &  0.07(0.05) &  0.08(0.05)  &      - &  8245\\%*   \\ 
1996X & 13.26(0.04)  &  1.09(0.07) &  1.32(0.05) & 0.069 & -0.00(0.06) &  0.07(0.07) &  0.03(0.05)  & 32.29 &  2325   \\ 
1996ai & 16.96(0.20)  &  0.83(0.02) &  0.85(0.08) & 0.014 &  1.76(0.06) &  2.07(0.10) &  1.88(0.22)  & 31.64 &  1177   \\ 
1996bk & 14.84(0.20)  &     -    &  1.69(0.20) & 0.018 &  0.38(0.21) &  0.33(0.13) &  0.35(0.05)  & 32.55 &  2147   \\ 
1996bl & 17.08(0.04)  &  0.99(0.03) &  1.11(0.05) & 0.092 &  0.05(0.06) &  0.13(0.07) &  0.09(0.05)  &      - & 10447\\%*   \\ 
1996bo & 16.15(0.04)  &  1.06(0.07) &  1.30(0.05) & 0.077 &  0.35(0.06) &  0.30(0.10) &  0.33(0.05)  &      - &  4898   \\ 
1996bv & 15.77(0.04)  &     -    &  0.84(0.10) & 0.105 &  0.23(0.20) &  0.23(0.10) &  0.23(0.05)  &      - &  5016   \\ 
1997bp & 14.10(0.20)  &  0.95(0.09) &  0.94(0.20) & 0.044 &  0.09(0.20) &  0.45(0.20) &  0.27(0.26)  &      - &  2824   \\ 
1997br & 14.06(0.04)  &  1.04(0.07) &  1.04(0.05) & 0.113 &  0.31(0.20) &  0.33(0.04) &  0.33(0.05)  &      - &  2399   \\ 
1998bu & 12.20(0.04)  &  1.00(0.06) &  1.15(0.05) & 0.025 &  0.39(0.06) &  0.36(0.05) &  0.37(0.05)  & 29.54 &  1238   \\ 
1998de & 17.56(0.02)  &  1.45(0.13) &  1.94(0.05) & 0.059 & -0.08(0.07) &  0.00(0.10) &  0.00(0.05)  &      - &  4713   \\ 
1999aa & 14.93(0.05)  &  0.85(0.07) &  0.85(0.05) & 0.040 &  0.11(0.06) &  0.07(0.05) &  0.09(0.05)  &      - &  4564   \\ 
1999aw & 16.86(0.05)  &  0.74(0.05) &  0.81(0.05) & 0.032 &  0.10(0.09) &  0.00(0.10) &  0.05(0.07)  &      - & 11362\\%*   \\ 
1999da & 16.90(0.10)  &     -    &  1.94(0.20) & 0.058 &      -    &  0.05(0.20) &  0.05(0.20)  &      - &  3644   \\ 
1999dk & 15.04(0.05)  &  0.92(0.08) &  1.28(0.10) & 0.054 &  0.13(0.06) &  0.17(0.21) &  0.14(0.05)  &      - &  4184   \\ 
1999ee & 14.94(0.02)  &  0.86(0.04) &  0.92(0.05) & 0.020 &  0.36(0.06) &  0.39(0.06) &  0.38(0.05)  &      - &  3153   \\ 
1999gp & 16.25(0.05)  &  0.83(0.07) &  0.94(0.10) & 0.056 &  0.15(0.06) &  0.11(0.09) &  0.14(0.05)  &      - &  7811   \\ 
2000E & 14.31(0.05)  &  0.85(0.07) &  0.94(0.05) & 0.366 &  0.21(0.10) &  0.30(0.10) &  0.25(0.07)  & 31.91 &  1280   \\ 
2000bk & 16.98(0.10)  &  1.32(0.17) &  1.69(0.10) & 0.025 &      -    &  0.19(0.05) &  0.19(0.05)  &      - &  7976\\%*   \\ 
2000ce & 17.24(0.10)  &  0.86(0.07) &  0.94(0.10) & 0.057 &      -    &  0.65(0.14) &  0.65(0.14)  &      - &  4948   \\ 
2000cx & 13.44(0.05)  &  0.94(0.09) &  0.93(0.05) & 0.083 &  0.19(0.06) & -0.20(0.05) &  0.00(0.28)  & 32.53 &  2115   \\ 
2001el & 12.81(0.02)  &  1.02(0.06) &  1.15(0.05) & 0.014 &  0.13(0.06) &  0.33(0.05) &  0.24(0.14)  & 30.55 &  1091   \\ 
2002bo & 14.01(0.10)  &  1.02(0.06) &  1.17(0.05) & 0.025 &  0.48(0.06) &  0.47(0.05) &  0.47(0.05)  & 31.76 &  1641   \\ 
\noalign{\smallskip}
\hline
\end{tabular}
\end{center}
\end{table*}

\begin{table*}
\begin{center}
\caption{}
\label{tab22}
\begin{tabular}{l l | l l}
\hline
SN&  References & SN&  References \\
\noalign{\smallskip}
\hline
\hline 
1937C & B.E. Schaefer, 1994, ApJ, 426, 493                          &1983G & W. E. Harris et al., 1983, PASP, 95, 607         \\
      & M.J. Pierce \& G.H. Jacoby, 1995, AJ, 110, 2885             &      & R.J. Buta et al., 1985, PASP, 97, 229            \\
1972E & A. Ardeberg \& M. de Groot, 1973, A\&A, 28, 295             &      & S. Benetti et al., 1991, A\&A, 247, 410          \\
      & A.M. van Genderen, 1975, A\&A, 45, 429                      &      & D.Y. Tsvetkov, 1985, SA, 29, 211                  \\
      & T.A. Lee et al.,  1972, ApJ, 177, L59                       &1986G & M.M. Phillips et al., 1987, PASP, 99, 592    \\             
      & A.W.J. Cousins 1972, IBVS, 700, 1                           &      &  S. Cristiani et al., 1992, A\&A, 259, 63    \\             
1972J & F. Ciatti \& L. Rosino, 1977, A\&A 57, 73 *                 &      & B.E. Schaefer 1987, ApJ, 323, 47             \\             
      & * revised in F. Patat et al. 1997 A\&A, 317, 423            &1989B & R. Barbon et al., 1990, A\&A, 237, 79        \\             
1974G & F. Ciatti \& L. Rosino, 1977, A\&A 57, 73                   &      & L.A. Wells et al., 1994, AJ, 108, 2233       \\             
      & B. Patchett \& R. Wood,  1976, MNRAS, 175, 595              &1990N & P. Lira et al., 1998, AJ, 115, 234           \\             
      & I.D. Howarth,  1974, Mitt. Ver\"anderl Sterne, 6, 155       &1990O&  M. Hamuy et al., 1996, AJ, 112, 2408         \\             
1980N & M. Hamuy et al., 1991, AJ, 102, 208			    &1990T&  M. Hamuy et al., 1996, AJ, 112, 2408         \\             
      & P.F. Younger \& S. van den Bergh, 1985, A\&AS, 61, 365      &1990Y&  M. Hamuy et al., 1996, AJ, 112, 2408         \\             
1981B & R. Barbon; F. Ciatti; L. Rosino,  1982, A\&A, 116, 35 *     &1991S&  M. Hamuy et al., 1996, AJ, 112, 2408         \\
      & * revised in F. Patat et al. 1997 A\&A, 317, 423            &     &  ESO-Asiago SN Archive                        \\ 
      & R.J. Buta \& A. Turner, 1983, PASP, 95, 72  		    &1991T&  P. Lira et al.,  1998, AJ, 115, 234          \\
      & D.Y. Tsvetkov, 1982, SA, 8, 115  			    &     &  E. Cappellaro et al., 1997, A\&A, 328, 203   \\       
      & IAUC 3584 						    &     &  B. Schmidt et al.,  1994 ApJ, 434, 19        \\ 
1982B & D.Y. Tsvetkov, 1986, PZ, 22, 279 			    &     &  IAUC 5246, 5253, 5256, 5270, 5273, 5309      \\          
      & F. Ciatti et al., 1988, A\&A, 202, 15 		            &     &  W.B. Sparks et al., 1999, ApJ, 523, 585      \\          
      & R. Cadanau \& C., Trefzger, 1983, IBVS, 2382, 1 	    &1991U&  M. Hamuy et al., 1996, AJ, 112, 2408         \\
\noalign{\smallskip}					 	            
\hline								            
\end{tabular}							            
\end{center}							              
\end{table*}

\begin{table*}
\begin{center}
{\bf Table B2.}
\begin{tabular}{l l | l l}
\hline
SN&  References & SN&  References \\
\noalign{\smallskip}
\hline
\hline 
1991ag& M. Hamuy et al., 1996, AJ, 112, 2408& 		     1995D  & IAUC 6134,6149,6156,6166 \\        
1991bg&  B. Leibundgut et al., 1993, AJ, 105, 301&	             & K. Sadakane et al., 1996,  PASJ, 48, 51\\                  
      & A.V. Filippenko et al., 1992, AJ, 104, 1543 &	             & ESO-Asiago SN Archive\\                                    
      & M. Turatto et al., 1996, MNRAS, 283, 1& 	            &  A.G. Riess et al.,  1999, AJ, 117, 707\\                   
1992A&   N.B Suntzeff, 1996, in McCray Z. Wang (eds.), &     1995E &  A.G. Riess et al.,  1999, AJ, 117, 707\\                    
&\, Supernovae and Supernova Remnants,&			     1995ac&  A.G. Riess et al.,  1999, AJ, 117, 707\\                                                                                      
&\, p.41, Cambridge Univ. Press, Cambridge &		           &  ESO-Asiago SN Archive\\                                                                                                       
     &  E. Cappellaro et al., 1997, A\&A, 328, 203&	     1995ak&  A.G. Riess et al.,  1999, AJ, 117, 707\\                    
1992J&  M. Hamuy et al., 1996, AJ, 112, 2408&		     1995al&  IAUC 6255,6256\\                                            
1992K&  ESO-Asiago SN Archive&				           &  A.G. Riess et al.,  1999, AJ, 117, 707\\                       
     &  M. Hamuy et al., 1996, AJ, 112, 2408&		           &  D. Yu. Tsevtkov et al. 2001 Astr. Rep. 45, 527\\              
1992P&  M. Hamuy et al., 1996, AJ, 112, 2408&		     1995bd&  A.G. Riess et al.,  1999, AJ, 117, 707\\                      
1992ae& M. Hamuy et al., 1996, AJ, 112, 2408&		           &  ESO-Asiago SN Archive\\                                       
1992ag& M. Hamuy et al., 1996, AJ, 112, 2408&		     1996C &  A.G. Riess et al.,  1999, AJ, 117, 707\\                      
 1992al& M. Hamuy et al., 1996, AJ, 112, 2408&       	          1996X &  IAUC 6380,6381\\                                      
 1992aq& M. Hamuy et al., 1996, AJ, 112, 2408&                          &  M.E. Salvo et al. 2001, MNRAS  321, 254\\             
 1992au& M. Hamuy et al., 1996, AJ, 112, 2408&       	                &  A.G. Riess et al.,  1999, AJ, 117, 707\\              
 1992bc& M. Hamuy et al., 1996, AJ, 112, 2408&       	          1996ai&  A.G. Riess et al.,  1999, AJ, 117, 707\\              
1992bg& M. Hamuy et al., 1996, AJ, 112, 2408&        	          1996bk&  A.G. Riess et al.,  1999, AJ, 117, 707\\              
1992bh& M. Hamuy et al., 1996, AJ, 112, 2408&        	          1996bl&  A.G. Riess et al.,  1999, AJ, 117, 707\\              
1992bk& M. Hamuy et al., 1996, AJ, 112, 2408&        	          1996bo&  A.G. Riess et al.,  1999, AJ, 117, 707\\              
1992bl& M. Hamuy et al., 1996, AJ, 112, 2408&        	                &  D. Yu. Tsevtkov et al. 2001 Astr. Rep. 45, 527\\      
1992bo& M. Hamuy et al., 1996, AJ, 112, 2408&        	                &  ESO-Asiago SN Archive\\                               
 1992bp& M. Hamuy et al., 1996, AJ, 112, 2408&       	          1996bv&  A.G. Riess et al.,  1999, AJ, 117, 707\\              
1992bs& M. Hamuy et al., 1996, AJ, 112, 2408&        	          1997bp&  ESO-Asiago SN Archive\\                               
1993B & M. Hamuy et al., 1996, AJ, 112, 2408&        	                &  A.G. Riess et al., 1998, ApJ, 504, 935\\              
 1993H & M. Hamuy et al., 1996, AJ, 112, 2408&       	          1997br&  W.D. Li et al., 1999, AJ, 117, 2709\\                 
         & IAUC 5723    &                            	                &  ESO-Asiago SN Archive\\                               
         & ESO-Asiago SN Archive&                    	          1998bu&  N.B. Suntzeff et al., 1999, AJ, 117, 1175\\           
1993L & IAUC 5780, 5781    &                         	                &  S. Jha et al. 1999,  ApJS, 125, 73\\                  
      & E. Cappellaro et al., 1997, A\&A, 328, 203&  	                &  E. Cappellaro et al., 2001, ApJ, 549, 215\\           
1993O & M. Hamuy et al., 1996, AJ, 112, 2408  &      	                &  M. Hernandez et al., 2000, MNRAS, 319, 223\\          
1993ac& A.G. Riess et al., 1999,  AJ, 117, 707&      	          1998de&  M. Modjaz et al., 2001, PASP, 113, 308\\              
1993ae& A.G. Riess et al., 1999,  AJ, 117, 707 &     	         1999aa&  ESO-Asiago SN Archive\\                                
      & W.C.G. Ho et al., 2001, PASP, 113, 1349& 	               &  K. Krisciunas et al., 2000, ApJ, 539 ,658\\            
1993ag& M. Hamuy et al., 1996, AJ, 112, 2408&                    1999aw&  L.G. Strolger et al.,  2002, AJ, 124, 2905\\          
1993ah& M. Hamuy et al., 1996, AJ, 112, 2408&                    1999da&  K. Krisciunas et al., 2001, AJ, 122, 1616\\           
1994D & F. Patat et al., 1996, MNRAS, 278, 111&                  1999dk&  K. Krisciunas et al., 2001, AJ, 122, 1616\\           
    & E. Cappellaro et al., 1997, A\&A, 328, 203&                      &  ESO-Asiago SN Archive\\                               
    & S. D. Van Dyk et al., 1999, AJ, 118, 2331&                 1999ee&  M.D.  Stritzinger  et al.,  2002, AJ, 124, 2100 \\    
    & M.W. Richmond et al., 1995, AJ 109, 2121&                  1999gp&  K. Krisciunas et al., 2001, AJ, 122, 1616\\           
    & W.P.S. Meikle et al., 1996, MNRAS, 281, 263&               2000E &  G. Valentini et al. astro-ph/0306391\\                
    & D.Y. Tsvetkov \& N.N. Pavlyuk, 1995, AstL, 21, 606&               & J. Vink\'o et al., 2001,  A\&A, 372, 824\\            
    & H. Wu, H.J. Yan, Z.L. Zou, 1995, A\&A, 294, 9&             2000bk&  K. Krisciunas et al., 2001, AJ, 122, 1616\\           
         & M.W. Richmond et al., 1995, AJ, 109, 2121&            2000ce&  K. Krisciunas et al., 2001, AJ, 122, 1616\\           
1994M & A.G. Riess et al., 1999,  AJ, 117, 707 &                 2000cx&   P. Candia et al., 2003, PASP, 115, 277 \\            
         &  W.C.G. Ho et al., 2001, PASP, 113, 1349&                   &  ESO-Asiago SN Archive\\                               
         & ESO-Asiago SN Archive&                                2001el&  K. Krisciunas et al., 2003, AJ, 125, 166 \\           
1994Q & A.G. Riess et al., 1999,  AJ, 117, 707 &                 2002bo&  IAUC 7863\\                                           
         &  W.C.G. Ho et al., 2001, PASP, 113, 1349&                   &  S. Benetti et al., 2003, MNRAS accepted\\             
1994T & A.G. Riess et al., 1999,  AJ, 117, 707 &                       & Gy.M. Szabo et al., 2003, A\&A, 408, 915\\             
1994ae & IAUC 6105,6111&                       & \\                  
    & A.G. Riess et al.,  1999, AJ, 117, 707&    & \\                
           & ESO-Asiago SN Archive               & & \\                       
           &  W.C.G. Ho et al., 2001, PASP, 113, 1349 & &  \\    
            & D.Y. Tsvetkov et al., 1997, AstL, 23, 26  & & \\  
\noalign{\smallskip}					 	            
\hline								            
\end{tabular}							            
\end{center}							              
\end{table*}

%%%%%%%%%%%%%%%%%%%%%%%%%%%%%%%%%%%%%%%%%%%%%%%%%%%%%%%%%%%%%%%%%%%%%%%%%%%%%%%
\clearpage
\newpage
\section{B band templates}\label{templateBnewsec}

\floatplacement{table}{h}
\begin{table*}
\begin{center}
\caption{ Light curve templates for different decline rates.}
\label{templateBnew}
\begin{tabular}[b]{l c c c c c c c c | l c c c c c c c}
\hline
\noalign{\smallskip}
$\Delta m_{15}$ & 0.82 & 1.02 & 1.11 & 1.29 & 1.42 & 1.67 & 1.94  &  & $\Delta m_{15}$   & 0.82 & 1.02 & 1.11 & 1.29 & 1.42 & 1.67 & 1.94 \\
\noalign{\smallskip}
\hline
\noalign{\smallskip}
phase  & \multicolumn{7}{c}{$m(B)$} & & phase  & \multicolumn{7}{c}{$m(B)$} \\
\noalign{\smallskip}
\hline
\hline
\noalign{\smallskip}
%  -15  & 2.031 & 2.730 & 1.720 & 1.833 & 3.168 & 1.771  & 3.882 &    
%  -14  & 1.840 & 2.403 & 1.510 & 1.690 & 2.828 & 1.630  & 3.562 &    
%  -13  & 1.649 & 2.077 & 1.300 & 1.547 & 2.487 & 1.489  & 3.242 &    
%  -12  & 1.457 & 1.750 & 1.091 & 1.404 & 2.147 & 1.348  & 2.922 &    
%  -11  & 1.266 & 1.424 & 0.884 & 1.261 & 1.807 & 1.206  & 2.602 &  
%  -10  & 1.075 & 1.095 & 0.687 & 1.118 & 1.466 & 1.065  & 2.282 &  
%   -9  & 0.884 & 0.793 & 0.509 & 0.975 & 1.132 & 0.924  & 1.962 &  
%   -8  & 0.693 & 0.541 & 0.356 & 0.832 & 0.826 & 0.783  & 1.642 &  
   -7  & 0.510 & 0.361 & 0.238 & 0.689 & 0.578 & 0.641  & 1.322 &   & 37  & 2.745 & 2.759 & 2.963 & 3.012 & 3.155 & 3.120  & 2.670\\ 
   -6  & 0.347 & 0.246 & 0.155 & 0.546 & 0.396 & 0.500  & 1.002 &   & 38  & 2.791 & 2.802 & 2.991 & 3.048 & 3.180 & 3.141  & 2.695\\ 
   -5  & 0.217 & 0.168 & 0.098 & 0.405 & 0.264 & 0.363  & 0.683 &   & 39  & 2.833 & 2.843 & 3.016 & 3.083 & 3.202 & 3.162  & 2.722\\ 
   -4  & 0.125 & 0.107 & 0.062 & 0.274 & 0.166 & 0.238  & 0.400 &   & 40  & 2.873 & 2.883 & 3.036 & 3.118 & 3.223 & 3.181  & 2.750\\ 
   -3  & 0.064 & 0.057 & 0.038 & 0.161 & 0.090 & 0.134  & 0.229 &   & 41  & 2.909 & 2.920 & 3.055 & 3.150 & 3.243 & 3.200  & 2.778\\ 
   -2  & 0.025 & 0.021 & 0.019 & 0.073 & 0.038 & 0.059  & 0.125 &   & 42  & 2.942 & 2.956 & 3.071 & 3.180 & 3.261 & 3.217  & 2.806\\ 
   -1  & 0.005 & 0.002 & 0.005 & 0.019 & 0.008 & 0.015  & 0.028 &   & 43  & 2.973 & 2.989 & 3.086 & 3.205 & 3.277 & 3.234  & 2.832\\ 
    0  & 0.000 & 0.000 & 0.000 & 0.000 & 0.000 & 0.000  & 0.000 &   & 44  & 3.001 & 3.019 & 3.099 & 3.225 & 3.291 & 3.251  & 2.856\\ 
    1  & 0.008 & 0.015 & 0.011 & 0.010 & 0.014 & 0.014  & 0.036 &   & 45  & 3.027 & 3.047 & 3.113 & 3.240 & 3.304 & 3.267  & 2.877\\ 
    2  & 0.027 & 0.041 & 0.042 & 0.040 & 0.048 & 0.053  & 0.106 &   & 46  & 3.052 & 3.073 & 3.127 & 3.254 & 3.315 & 3.283  & 2.896\\ 
    3  & 0.054 & 0.073 & 0.085 & 0.084 & 0.101 & 0.114  & 0.186 &   & 47  & 3.075 & 3.096 & 3.142 & 3.266 & 3.324 & 3.299  & 2.914\\ 
    4  & 0.087 & 0.107 & 0.132 & 0.140 & 0.168 & 0.191  & 0.296 &   & 48  & 3.095 & 3.116 & 3.157 & 3.279 & 3.331 & 3.315  & 2.930\\ 
    5  & 0.124 & 0.148 & 0.181 & 0.207 & 0.246 & 0.280  & 0.465 &   & 49  & 3.116 & 3.135 & 3.173 & 3.294 & 3.339 & 3.330  & 2.946\\ 
    6  & 0.166 & 0.199 & 0.236 & 0.285 & 0.331 & 0.377  & 0.647 &   & 50  & 3.135 & 3.151 & 3.189 & 3.311 & 3.351 & 3.346  & 2.963\\ 
    7  & 0.213 & 0.262 & 0.300 & 0.371 & 0.421 & 0.485  & 0.844 &   & 51  & 3.154 & 3.166 & 3.204 & 3.329 & 3.364 & 3.361  & 2.982\\ 
    8  & 0.268 & 0.337 & 0.378 & 0.466 & 0.518 & 0.605  & 1.046 &   & 52  & 3.172 & 3.180 & 3.220 & 3.349 & 3.382 & 3.377  & 3.001\\ 
    9  & 0.332 & 0.421 & 0.468 & 0.569 & 0.627 & 0.741  & 1.240 &   & 53  & 3.188 & 3.193 & 3.235 & 3.369 & 3.402 & 3.393  & 3.021\\ 
   10  & 0.402 & 0.509 & 0.565 & 0.680 & 0.745 & 0.891  & 1.417 &   & 54  & 3.203 & 3.205 & 3.250 & 3.388 & 3.423 & 3.409  & 3.042\\ 
   11  & 0.477 & 0.601 & 0.666 & 0.798 & 0.873 & 1.049  & 1.565 &   & 55  & 3.219 & 3.216 & 3.265 & 3.408 & 3.444 & 3.426  & 3.062\\ 
   12  & 0.554 & 0.698 & 0.766 & 0.924 & 1.010 & 1.211  & 1.687 &   & 56  & 3.233 & 3.227 & 3.279 & 3.427 & 3.465 & 3.443  & 3.082\\ 
   13  & 0.634 & 0.802 & 0.871 & 1.051 & 1.150 & 1.369  & 1.787 &   & 57  & 3.247 & 3.238 & 3.294 & 3.445 & 3.485 & 3.460  & 3.102\\ 
   14  & 0.721 & 0.910 & 0.985 & 1.173 & 1.288 & 1.521  & 1.868 &   & 58  & 3.261 & 3.249 & 3.307 & 3.463 & 3.504 & 3.477  & 3.122\\ 
   15  & 0.815 & 1.018 & 1.109 & 1.285 & 1.418 & 1.664  & 1.935 &   & 59  & 3.274 & 3.260 & 3.321 & 3.481 & 3.523 & 3.495  & 3.142\\ 
   16  & 0.915 & 1.122 & 1.239 & 1.390 & 1.539 & 1.799  & 1.992 &   & 60  & 3.288 & 3.272 & 3.334 & 3.497 & 3.540 & 3.513  & 3.161\\ 
   17  & 1.019 & 1.225 & 1.371 & 1.491 & 1.656 & 1.926  & 2.042 &   & 61  & 3.301 & 3.284 & 3.346 & 3.513 & 3.557 & 3.531  & 3.181\\ 
   18  & 1.126 & 1.327 & 1.499 & 1.593 & 1.769 & 2.046  & 2.088 &   & 62  & 3.314 & 3.298 & 3.359 & 3.528 & 3.574 & 3.550  & 3.201\\ 
   19  & 1.235 & 1.428 & 1.622 & 1.697 & 1.880 & 2.158  & 2.133 &   & 63  & 3.327 & 3.312 & 3.371 & 3.542 & 3.590 & 3.569  & 3.221\\ 
   20  & 1.344 & 1.529 & 1.739 & 1.802 & 1.990 & 2.262  & 2.181 &   & 64  & 3.340 & 3.327 & 3.383 & 3.555 & 3.605 & 3.588  & 3.242\\ 
   21  & 1.453 & 1.629 & 1.851 & 1.908 & 2.099 & 2.359  & 2.231 &   & 65  & 3.353 & 3.342 & 3.395 & 3.567 & 3.620 & 3.607  & 3.263\\ 
   22  & 1.560 & 1.727 & 1.958 & 2.013 & 2.205 & 2.447  & 2.280 &   & 66  & 3.366 & 3.358 & 3.406 & 3.578 & 3.636 & 3.627  & 3.284\\ 
   23  & 1.665 & 1.822 & 2.059 & 2.116 & 2.308 & 2.529  & 2.328 &   & 67  & 3.379 & 3.374 & 3.417 & 3.589 & 3.650 & 3.646  & 3.306\\ 
   24  & 1.769 & 1.915 & 2.156 & 2.217 & 2.408 & 2.602  & 2.372 &   & 68  & 3.391 & 3.391 & 3.428 & 3.600 & 3.664 & 3.666  & 3.329\\ 
   25  & 1.869 & 2.005 & 2.247 & 2.314 & 2.504 & 2.668  & 2.413 &   & 69  & 3.404 & 3.409 & 3.439 & 3.611 & 3.679 & 3.685  & 3.353\\ 
   26  & 1.966 & 2.091 & 2.333 & 2.406 & 2.595 & 2.729  & 2.449 &   & 70  & 3.417 & 3.427 & 3.449 & 3.621 & 3.693 & 3.705  & 3.378\\ 
   27  & 2.059 & 2.172 & 2.414 & 2.493 & 2.680 & 2.782  & 2.480 &   & 71  & 3.430 & 3.445 & 3.459 & 3.632 & 3.707 & 3.724  & 3.404\\ 
   28  & 2.147 & 2.248 & 2.491 & 2.572 & 2.758 & 2.831  & 2.507 &   & 72  & 3.442 & 3.463 & 3.469 & 3.643 & 3.720 & 3.743  & 3.430\\ 
   29  & 2.230 & 2.321 & 2.562 & 2.644 & 2.828 & 2.874  & 2.531 &   & 73  & 3.455 & 3.481 & 3.479 & 3.655 & 3.734 & 3.762  & 3.458\\ 
   30  & 2.310 & 2.387 & 2.629 & 2.709 & 2.889 & 2.914  & 2.550 &   & 74  & 3.468 & 3.500 & 3.489 & 3.667 & 3.749 & 3.780  & 3.488\\ 
   31  & 2.385 & 2.450 & 2.691 & 2.767 & 2.944 & 2.951  & 2.567 &   & 75  & 3.481 & 3.519 & 3.498 & 3.680 & 3.763 & 3.798  & 3.518\\ 
   32  & 2.455 & 2.509 & 2.749 & 2.819 & 2.990 & 2.984  & 2.582 &   & 76  & 3.493 & 3.537 & 3.507 & 3.694 & 3.778 & 3.816  & 3.549\\ 
   33  & 2.521 & 2.565 & 2.801 & 2.864 & 3.031 & 3.016  & 2.597 &   & 77  & 3.506 & 3.556 & 3.516 & 3.708 & 3.794 & 3.833  & 3.581\\    
   34  & 2.584 & 2.617 & 2.849 & 2.905 & 3.067 & 3.044  & 2.612 &   & 78  & 3.519 & 3.574 & 3.525 & 3.723 & 3.809 & 3.849  & 3.613\\
   35  & 2.641 & 2.667 & 2.892 & 2.942 & 3.099 & 3.071  & 2.628 &   & 79  & 3.532 & 3.593 & 3.534 & 3.739 & 3.825 & 3.865  & 3.644\\
   36  & 2.695 & 2.714 & 2.930 & 2.977 & 3.128 & 3.096  & 2.648 &   & 80  & 3.545 & 3.611 & 3.542 & 3.756 & 3.842 & 3.880  & 3.676\\
\noalign{\smallskip}
\hline	
\end{tabular}				     
\end{center}							      
\end{table*}

\end{document}